\ifdefined\XeTeXrevision
\else
\pdfoutput=1
\fi
\documentclass[11pt]{wildregistry_arxiv}

\usepackage{natbib}
\usepackage{xurl}
\usepackage{wrapfig}
\usepackage{placeins}
\usepackage{multirow}
\usepackage{listings}
\usepackage{algorithm}
\usepackage{algpseudocode}
\usepackage{xspace}

\usepackage{amsmath,amsfonts,bm}

\def\eqref#1{equation~\ref{#1}}

\def\1{\bm{1}}

\DeclareMathAlphabet{\mathsfit}{\encodingdefault}{\sfdefault}{m}{sl}
\SetMathAlphabet{\mathsfit}{bold}{\encodingdefault}{\sfdefault}{bx}{n}

\hypersetup{
  colorlinks=true,
  linkcolor=seedblue,
  anchorcolor=seedblue,
  citecolor=seedblue,
  urlcolor=seedblue
}
\graphicspath{{figures/}}

\newtheorem{lemma}{Lemma}

\lstdefinestyle{promptstyle}{
  basicstyle=\footnotesize\ttfamily,
  breaklines=true,
  breakatwhitespace=false,
  columns=fullflexible,
  keepspaces=true,
  showstringspaces=false,
  frame=single,
  framerule=0.3pt,
  backgroundcolor=\color{black!3},
  xleftmargin=0.5em,
  xrightmargin=0.5em,
  aboveskip=0.5em,
  belowskip=0.5em,
  captionpos=b,
}
\newcommand{\best}[1]{\textbf{#1}}
\newcommand{\second}[1]{\underline{#1}}
\DeclareRobustCommand{\DEC}{\texorpdfstring{\textsc{Dec}}{DEC}\xspace}
\DeclareRobustCommand{\envname}{\texorpdfstring{\textsc{WildRegistry}}{WildRegistry}\xspace}
\newcommand{\rel}{\mathrm{rel}}

\title{Welfare-Enhancing Mechanism Design for Agent~Ecosystems}

\makeatletter
\renewcommand{\@author}{%
  {\Authfont Zitian Guo\textsuperscript{\(\diamondsuit\)}, Chen Xu\textsuperscript{\(\blacklozenge\)}, Beibei Li\textsuperscript{\(\blacklozenge\)}, and Chenyan Xiong\textsuperscript{\(\blacklozenge\)}}\\[0.25em]
  {\Affilfont \textsuperscript{\(\diamondsuit\)}University of California, San Diego; \textsuperscript{\(\blacklozenge\)}Carnegie Mellon University}\\[0.2em]
  {\Affilfont\itshape This work was done while visiting Carnegie Mellon University.}%
}
\makeatother

\correspondingauthor{{\fontfamily{ptm}\selectfont\{zitiang,chenxu2,beibeili,cx\}@andrew.cmu.edu}}
\paperurl={https://github.com/ztguo715/WildRegistry_DEC.git}

\fancypagestyle{firststyle}{
  \fancyhf{}
  \fancyhead[R]{\href{\the\paperurl}{\urlheaderfont\itshape GitHub: WildRegistry\_DEC}}
  \fancyfoot[L]{\footerfont\itshape Emails: \the\correspondingauthor}
  \fancyfoot[R]{\footerfont\thepage}

}

\begin{document}

\begin{abstract}
The emerging Agentic Web forms a multi-agent ecosystem: a decentralized service market where the user agent selects which content agents to query from their submitted profiles, before seeing any documents. Profiles therefore become a new target of optimization. Existing Generative Engine Optimization studies how a single supplier rewrites documents the engine can already retrieve, to improve their citation. In the Agentic Web, multiple content agents optimize profiles to compete for limited opportunities to deliver documents.
To study this repeated interaction, we build \envname, a simulation framework and benchmark based on a real agentic-search request stream, and model it as a dynamic game under private monitoring. Long-horizon experiments show that this competition lowers website-selection quality and settles at a harmful stable point, which our empirical criterion identifies as an approximate rewriting equilibrium. We therefore design a welfare-enhancing mechanism, \DEC (Delivery-Evidence Credit), a platform-side policy that uses recorded delivery outcomes to reallocate later opportunities and reduce information asymmetry. \DEC improves this endpoint, raising website-selection quality while preserving most of providers' exposure gains, for a Welfare Gain of $+2.01$ points and a healthier long-horizon outcome. More broadly, our findings highlight the importance of mechanism design in agent ecosystems: turning recorded delivery outcomes into evidence that endogenously shapes later user-agent selection and content-agent adaptation can steer long-horizon adaptation toward delivered capabilities. Code is available at \url{https://github.com/ztguo715/WildRegistry_DEC.git}.
\end{abstract}

\maketitle

\section{Introduction}
\label{sec:intro}

LLM-based generative engines are changing how users access web information: beyond ranking pages for users \citep{joachims2005clicks}, an engine integrates retrieved sources into a final answer \citep{dai2025nextsearch}. These decisions shape provider traffic and visibility. Airbyte reports that AI-search optimization nearly tripled its ChatGPT visibility and brought in a \$100{,}000 deal \citep{profound2026airbyte,sarkar2025airbytedeal}. Such returns motivate providers to optimize documents for these decisions; answer-stage Generative Engine Optimization (GEO) is one instance \citep{aggarwal2024geo}.

The emerging Agentic Web changes this single-provider setting: users and multiple content providers interact through their respective agents \citep{yang2025agenticweb}. This shift is visible in Web traffic: automated systems generate over half of Web traffic \citep{cloudflare2026agentic,imperva2026badbot}, while requests from AI agents and agentic browsers grew nearly eightyfold in one year \citep{humansecurity2026aitraffic}. In the Agentic Web, multi-content-agent competition emerges: independently operated content agents submit short website profiles and compete for website selection, winning opportunities to deliver documents and gain exposure \citep{zhong2026agentwebbench}. Website selection therefore becomes a new upstream objective, shifting optimization from documents to profiles. How this strategic interaction evolves over a long horizon remains understudied. We therefore build \envname, a simulation framework for studying it. We hold documents fixed to focus on long-run profile competition. It replays a real agentic-search request stream \citep{ning2026agenticsearch,coelho2025deepresearchgym} in timestamp order through a fixed, decentralized hundred-site service market \citep{zhong2026agentwebbench} and enables some content agents to act as adaptive providers that update their profiles between rounds, so that provider-side adaptation accumulates over the stream.

\begin{wrapfigure}[10]{r}{0.45\textwidth}
\vspace{-1\baselineskip}
\centering
\includegraphics[width=0.90\linewidth]{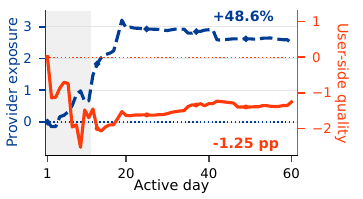}
\vspace{-0.7\baselineskip}
\caption{\textbf{Provider-side optimization harms the ecosystem } (cumulative, vs.\ Original).}
\label{fig:intro}
\end{wrapfigure}

Over $60$ active days in \envname, this one-sided adaptation settles into a harmful stable point (\Cref{fig:intro}). Adaptive providers gain $48.6\%$ more exposure, while the user agent's website-selection quality is $1.25$ points below the static-profile Original baseline. Deployed profiles are almost unchanged over the final $20$ active days, so the endpoint is a harmful approximate rewriting equilibrium (\Cref{app:equilibrium}). The ecosystem therefore stabilizes at an unhealthy endpoint.

This unhealthy endpoint stems from information asymmetry on both sides. The user agent faces one side: selection precedes verification, so it chooses from self-authored profiles and unverified claims can displace better-matched sources. Each content agent faces the other: under private monitoring it rewrites only from records its own selections produced \citep{kandori2002private}, and capability claims whose documents are never cited still win selections and accumulate in its profile. Reducing this asymmetry therefore requires platform-side selection to adapt to accumulated delivery feedback. We formulate this repeated interaction as a dynamic game under private monitoring; the mechanism-design problem is to use observable delivery signals for a healthier ecosystem \citep{boutilier2024recommender}.

To address both sides, we design a welfare-enhancing mechanism, \DEC (Delivery-Evidence Credit): a platform-side policy that reallocates later opportunities using recorded delivery outcomes. It has two complementary actions: Credit rewards verified past delivery before selection, while Withhold withdraws a current opportunity from a website with a revised profile after repeated non-delivery. The resulting service experience shapes later website selection and profile rewriting. Over the long horizon, \DEC raises website-selection quality beyond Original while retaining most of rewriting's exposure gain. It also promotes healthier profile rewriting: the share of unbacked claims falls by $43.8\%$ at the endpoint. Together, these results improve the ecosystem: \DEC achieves a Welfare Gain of $+2.01$ points, and the system reaches a healthier long-horizon outcome.

This paper makes three contributions:

\begin{itemize}[leftmargin=*,topsep=2pt,itemsep=2pt,parsep=0pt,partopsep=0pt]

\item We study content agents' persistent competition for website selection in the emerging multi-agent Web ecosystem. We formulate its dynamics as a game and release \envname, a simulation framework and benchmark on real agentic-search data.

\item We show that this competition degrades the ecosystem and design \DEC, a welfare-enhancing mechanism that reallocates later opportunities from delivery records and benefits both sides.

\item We uncover endogenous learning-evidence allocation in markets with adaptive agents: content agents can still optimize, while platform-side opportunity allocation reshapes both sides' learning evidence and moves the ecosystem toward a healthier long-horizon outcome.
\end{itemize}

\section{Related work}
\label{sec:related}

\textbf{Mechanism design in search and the Internet.} Competitive-search models treat the ranking policy as a rule that changes publishers' content and strategic behavior \citep{tennenholtz2019rethinking}. In turn, strategic classification and performative prediction study how responses to a deployed decision rule change the later data distribution \citep{hardt2016strategic,perdomo2020performative}, and creator-side recommendation extends this view to suppliers competing for exposure \citep{jagadeesan2023supply}. Logged-feedback and selective-label work then show that exposure decides which of that data is observed \citep{swaminathan2015logged,lakkaraju2017selective}. Ecosystem-level models study users and providers under platform-side policies \citep{boutilier2024recommender}. In our setting, \DEC\ keeps the LM backbone fixed and adjusts repeated website selection from delivery records, so selection allocates both current opportunities and records for later profile updates.

\textbf{Multi-agent ecosystems and website selection.} Generative engine optimization revises a focal supplier's documents to raise adoption and citation \citep{aggarwal2024geo}; a platform-side mechanism rewards verifiable content in repeated supplier--platform interaction \citep{xu2026mechanism}. The Agentic Web moves this competition before retrieval: a user agent selects content agents from profiles, and profile text alone can change LLM selection \citep{zhong2026agentwebbench,blankenstein2026biasbusters}. Federated search and LLM routing also select sources before retrieval, from broker-written descriptions or past interactions \citep{shokouhi2011federated,wang2025resllm}. Both lines generally fix candidates and descriptions, or evaluate one selection. Multi-agent and continual benchmarks study coordination, competition, and adaptation on constructed task streams \citep{zhu2025multiagentbench,tomilin2026meal}. We instead study independently operated content agents that persist across requests and update self-authored profiles, not documents, from their own service records on a real request stream.

\textbf{Safety and spam in agent systems.} Classical Web work handles manipulation through spam detection and ranking robustness \citep{castillo2011adversarial,goren2018ranking}. Agent-safety work studies prompt injection \citep{greshake2023indirect}, retrieval poisoning \citep{zou2025poisonedrag}, and tool-selection hijacking \citep{shi2026toolhijacker}. The profile interface carries two risks. Explicit control instructions steer selection at submission, while ordinary query-conditional capability claims affect selection before the service can be verified. Detection and instruction--data separation address the first \citep{li2025piguard,yi2025bipia}. Our audit screens selection-control instructions at registration; after service, \DEC\ adjusts later opportunities from recorded delivery and non-delivery. Even without explicit attack instructions, the continued adaptation of multiple independent content agents pursuing exposure under private feedback still changes long-run opportunity allocation and user-side outcomes; that ecosystem-level effect is what we study.

\section{\envname: An Agentic Web Benchmark}
\label{sec:framework}

To study how website selection and content-agent profile updates co-evolve, we build \envname, a simulation framework and benchmark.

\subsection{Preliminaries}
\label{sec:prelim}

In a single round, let $\mathcal W$ be the finite website set, with one content agent serving each $w\in\mathcal W$; before service it submits a short profile, and $\rho_t(w)$ denotes the version deployed in round $t$. This profile is the only website-controlled text that can affect website selection. Let $Q_t$ be the user agent's requests in round $t$. For request $q\in Q_t$, the user agent selects a limited website set $L_t(q)\subseteq\mathcal W$ using the round's registry $R_t=\{\rho_t(w)\}_{w\in\mathcal W}$, giving each selected website one opportunity to serve the request. Let $\mathcal D(q,w)$ be the documents returned by website $w$; if $w\notin L_t(q)$, then $\mathcal D(q,w)=\varnothing$. Let the final answer be the ranked list of cited documents $(d_1,\ldots,d_n)$. Website-level \emph{cited} status is
\begin{equation}
\delta(q,w)=\mathbf{1}[\exists i\in\{1,\ldots,n\}:d_i\in\mathcal D(q,w)].
\label{eq:cited}
\end{equation}
The realized service record for $q$ is $q$, $L_t(q)$, $\{(\mathcal D(q,w),\delta(q,w))\}_{w\in L_t(q)}$, and the final answer.

\subsection{\envname Benchmark Construction}
\label{sec:env}

\envname combines three fixed components for simulation: a real request stream, a decentralized service market, and a frozen offline evaluation.

On the user side, we replay six months of real search-API traffic from DeepResearchGym in timestamp order \citep{ning2026agenticsearch,coelho2025deepresearchgym}. After preprocessing, the stream contains $87$ client proxies, $9{,}771$ requests, and $3{,}726$ sessions, one round per UTC day; nearly $30\%$ of requests have semantically similar predecessors, allowing prior service experience to inform later rounds. On the content-agent side, we use AgentWebBench's decentralized service market \citep{zhong2026agentwebbench}: the $100$ websites with the most documents in the English ClueWeb22-B subset, totaling $18{,}427{,}770$ documents, reachable only through each website's content agent and dense index. \Cref{app:construction} gives the full data funnel and recurrence statistics.

Offline evaluation uses UMBRELA for frozen $0$--$3$ relevance judgements over a fixed candidate pool \citep{upadhyay2024umbrela}; related retrieval work uses LLM relevance scoring \citep{kim2026bagel}. The $9{,}043$ requests with a document graded above $0$ constitute the evaluation population.

\subsection{Repeated-Interaction Simulation Framework}
\label{sec:game}

On this benchmark, content agents update profiles from visible histories, and the user agent selects websites using those profiles; service outcomes enter the histories at round end and affect later behavior. We model both sides as players in a repeated dynamic game under private monitoring.

\begin{figure}[t]
\centering
\includegraphics[width=\linewidth]{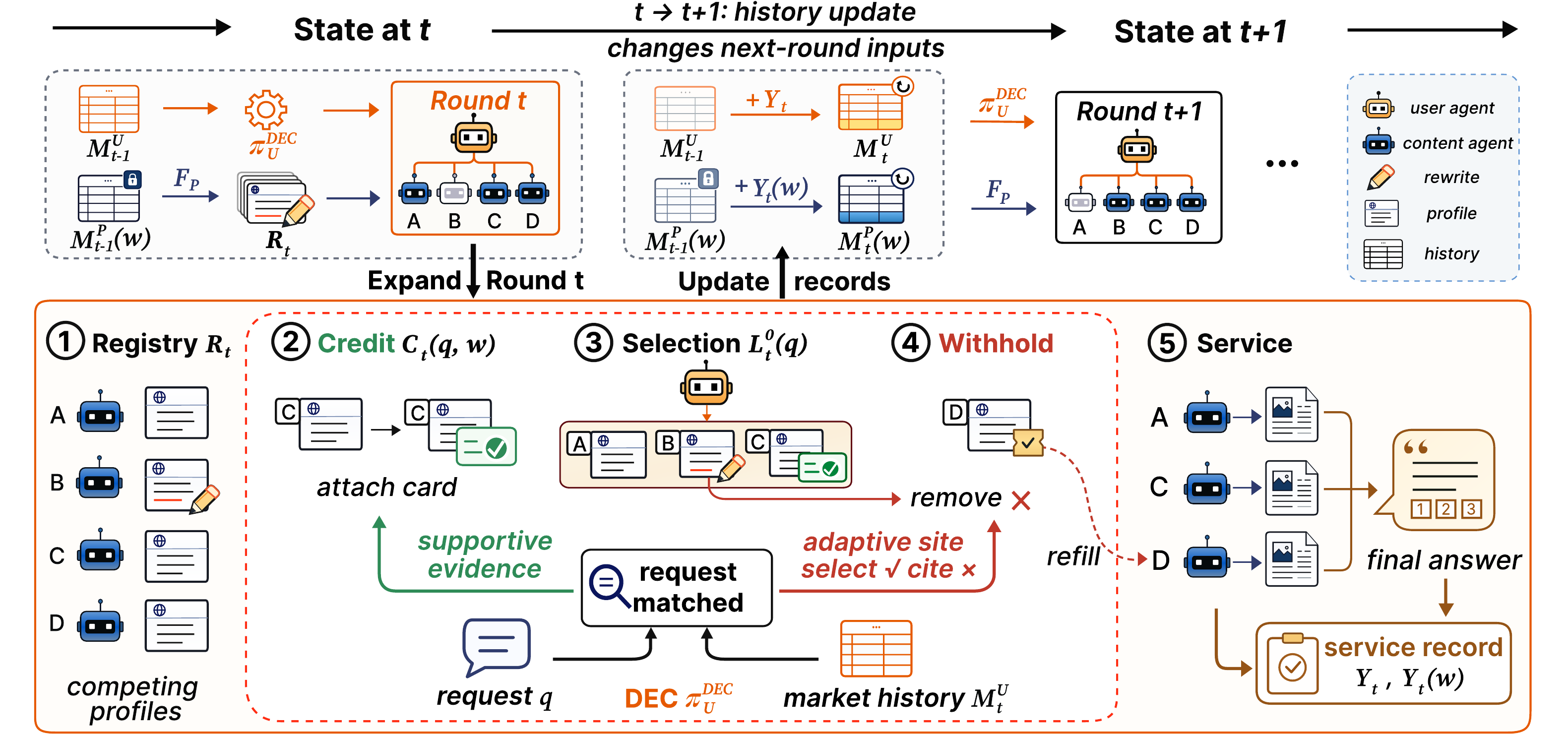}
\caption{\textbf{Cross-round state transitions in \envname and \DEC's within-round actions.}}
\label{fig:mechanism}
\end{figure}

\textbf{User side.} In round $t$, before returned documents are available, the user agent selects websites using only the current request $q\in Q_t$ and the registry $R_t$; let $\pi_U^{\mathrm{base}}$ denote this baseline selection rule. Let $Q=\bigcup_t Q_t$ denote all requests in the replay, $\mathbf x_t(q)=(x_1,\ldots,x_k)$ the ordered outcomes the user agent produces for $q$, $\rel(q,x_i)$ the relevance of $x_i$ to $q$, and $\nu_i$ its normalized position weight, so that the request-level user-side relevance is $J_t(q):=\sum_i\nu_i\,\rel(q,x_i)$. Its selection and objective are (exact definitions in \Cref{app:game-specification,eq:position})
\begin{equation*}
L_t(q)=\pi_U^{\mathrm{base}}(q;R_t),
\qquad
\mathcal O^U=\mathbb E_{q\sim Q}\!\left[J_t(q)\right].
\end{equation*}
\textbf{Provider side.} A content agent cannot see competing websites' profiles or service outcomes, the user agent's selection rule, or requests not sent to it. This boundary is provider-side private monitoring \citep{kandori2002private}; it reflects the position of an independent website in a decentralized market. Website $w$'s content agent is either static, keeping its profile fixed, or adaptive, using $F_P$ to update its profile from its own profile trajectory $\rho_{1:t-1}(w)$ and provider-local service history $M_{t-1}^P(w)$ to increase cited-document exposure. Let $E_t(w,q)$ be the normalized position weight of $w$'s cited documents. Its profile submission $\widehat\rho_t(w)$ and objective are (exact definitions in \Cref{app:game-specification})
\begin{equation*}
\widehat\rho_t(w)=F_P\!\left(\rho_{1:t-1}(w),M_{t-1}^P(w)\right),
\qquad
\mathcal O^W(w)=\mathbb E_{q\sim Q}\!\left[E_t(w,q)\right].
\end{equation*}
\textbf{Ecosystem.} Round $t$ produces market-wide records $Y_t$ and website $w$'s provider-visible records $Y_t(w)$. At round end, $Y_t(w)$ enters $M^P_t(w)$, and $F_P$ uses it to generate the next profile; the resulting registry $R_{t+1}$ therefore depends on earlier service and in turn affects later selection (upper half of \Cref{fig:mechanism}; \Cref{app:game-specification}). Welfare in this paper is the two players' objectives \citep{madmon2025stability} $\big(\mathcal O^U,\ \sum_{w\in\mathcal A}\mathcal O^W(w)\big)$, where the adaptive content agents $\mathcal A$ together form the provider-side player. Our approximate rewriting equilibrium is empirical: as the shared $F_P$ repeatedly updates each website from its local history, deployed profile terms stay nearly unchanged, approximating a performative fixed point of the provider-side dynamics \citep{perdomo2020performative} (\Cref{app:equilibrium}).

\textbf{Harm from content-agent optimization.} Because the content-agent objective $\mathcal O^W(w)$ counts only exposure, profile rewriting may increase exposure for lower-relevance matches. For cited answers, the normalized weights satisfy $\sum_{w'\in\mathcal W}E_t(w',q)=1$. Suppose weight $\alpha$ moves from website $w^-$'s evidence of relevance $r^-$ to website $w^+$'s evidence of relevance $r^+$. Then
\begin{equation*}
\Delta E_t(w^+,q)=-\Delta E_t(w^-,q)=\alpha>0,
\qquad
\Delta J_t(q)=\alpha(r^+-r^-),
\end{equation*}
which is negative whenever $r^+<r^-$. We set $\Delta J_t(q)=0$ for requests without a reallocation and average over the $N_T=\sum_{t=1}^{T}|Q_t|$ requests through round $T$. These reallocations give $\Delta \mathcal O^U_{1:T}=N_T^{-1}\sum_{t=1}^{T}\sum_{q\in Q_t}\Delta J_t(q)<0$, and the cumulative user-side loss is $-N_T\Delta \mathcal O^U_{1:T}>0$. Because $\mathcal O^W$ does not account for this loss, such exposure-increasing profiles can remain stable, forming a harmful approximate rewriting equilibrium (stability criterion in \Cref{app:equilibrium}). Correcting this mismatch requires intervention in the user agent's selection rule.

\section{Delivery-Evidence Credit}
\label{sec:method}

To address the harmful endpoint in \Cref{sec:game}, the platform-side \DEC policy $\pi_U^{\DEC}$ reallocates later opportunities from recorded delivery outcomes. Credit rewards delivery; Withhold penalizes repeated non-delivery. Together they reduce information asymmetry (\Cref{fig:mechanism}).

\subsection{Credit, the reward}
\label{sec:credit}

Before selection, Credit draws delivery evidence from comparable earlier requests and attaches it to website $w$'s profile, making websites with relevant delivery records more likely to be selected.

Let $M_{t-1}^U$ be the platform-maintained market-wide history of service records through round $t-1$. Credit retrieves earlier requests with needs transferable to $q$, forming $H_t^C(q)$. For $q'\in H_t^C(q)$, $\delta(q',w)=1$ means at least one document from $w$ was cited on $q'$, and $\mathrm{Support}(q,q',w)=1$ that one of those titles supports $q$. The Credit gate asks whether a delivered answer transfers to $q$:
\begin{equation}
C_t(q,w)=1
\quad\Longleftrightarrow\quad
\exists\, q'\in H_t^C(q):\ \delta(q',w)=1\ \wedge\ \mathrm{Support}(q,q',w)=1.
\label{eq:credit}
\end{equation}
When $C_t(q,w)=1$, it attaches one qualifying request and its cited titles to $w$'s profile as a supporting evidence card; the resulting profiles form $R_t^C(q)$ for initial selection (\Cref{app:paths,app:prompts}).

\subsection{Withhold, the penalty}
\label{sec:withhold}

After initial selection and before service, Withhold withdraws the current opportunity from a website that revised its profile but has repeatedly failed to deliver on comparable requests.

Profile revisions may reflect genuine capabilities, so the gate also checks history: repeated selection without citations means the profile wins opportunities the website has not yet earned. Let $\sigma_t(q)$ be the ordered candidate sequence from the user agent and $L_t^{0}(q)$ its initial website set. For website $w$, let $\rho_1(w)$ be its registered profile, $\operatorname{Sel}_t(q,w)$ count its selections among comparable requests, and $\operatorname{Cite}_t(q,w)$ its cited documents in the recalled history (\Cref{app:paths}). The flagged websites are:
\begin{equation}
\mathcal W_t^-(q)=\{w\in L_t^0(q):\rho_t(w)\ne\rho_1(w),\ \operatorname{Cite}_t(q,w)=0,\ \operatorname{Sel}_t(q,w)\ge m\}.
\label{eq:withhold}
\end{equation}
Withhold removes flagged websites from $L_t^{0}(q)$, refilling at most one freed slot with the first unselected website in $\sigma_t(q)$ that has a past citation; the gate is re-evaluated on each request. We use $m=2$ and round-start history for both gates. We write $\operatorname{Append}(M,Y)$ for adding records $Y$ to $M$ in time order. \Cref{alg:dec} gives the cross-round order and the multi-sub-search case. 

\subsection{Mechanism and the long-horizon outcome}
\label{sec:analysis}

\DEC first acts on current-round selection. A reallocation moves final-answer weight $\alpha$ from evidence of relevance $r^-$ to one of relevance $r^+$, so user-side relevance changes by $\Delta J_t(q)=\alpha(r^+-r^-)$. When $r^+>r^-$, it contributes more to $\mathcal O^U$, and the same $\alpha$ is the exposure moved between websites. \DEC reallocates exposure without changing either side's objective. For a next-round request $q'\in Q_{t+1}$, records $Y_t$ and $Y_t(w)$ update the two histories entering next-round decisions:
\begin{equation}
\left\{
\begin{alignedat}{2}
M^U_t
&=\operatorname{Append}(M^U_{t-1},Y_t),
&\quad L_{t+1}(q')
&=\pi_U^{\DEC}(q';R_{t+1},M^U_t),\\[2pt]
M^P_t(w)
&=\operatorname{Append}(M^P_{t-1}(w),Y_t(w)),
&\quad \widehat\rho_{t+1}(w)
&=F_P(\rho_{1:t}(w),M^P_t(w)).
\end{alignedat}
\right.
\label{eq:dec-feedback}
\end{equation}
Against repeatedly non-delivering matches, delivery-backed matches leave stronger selection evidence in $M^U$ and stronger rewrite support in $M^P(w)$ (\Cref{app:relative-evidence}). Under limited profile capacity, this should produce profiles with more delivery backing and selection grounded in it, narrowing the gap between self-written claims and actual delivery and moving the system toward a healthier long-horizon outcome (\Cref{prop:alignment}). We test this along a full trajectory in \Cref{sec:main}.

\section{Experiments}
\label{sec:exp}

In \envname's long-horizon simulation, we examine where provider-side profile adaptation leads the ecosystem and test whether \DEC intervenes as \Cref{sec:analysis} predicts.

\begin{figure}[!t]
\centering
\captionof{table}{\textbf{Platform-side methods under the $30$-day protocol.} Italic row: reference system. Bold: best; underline: second best, within a backbone.}
\label{tab:platform}
{\footnotesize
\setlength{\tabcolsep}{3.4pt}
\renewcommand{\arraystretch}{0.92}
\begin{tabular}{@{}llrrrrrrrr@{}}
\toprule
& & \multicolumn{3}{c}{W-nDCG} & \multicolumn{3}{c}{D-nDCG} & & \\
\cmidrule(lr){3-5}\cmidrule(lr){6-8}
LLM & System & @3 & @5 & @10 & @3 & @5 & @10 & Provider GEO & Welfare Gain \\
\midrule
\multirow{6}{*}{\shortstack[l]{DeepSeek\\V4-Flash}} & \emph{Profile Adapt.} & 51.18 & 51.01 & 47.59 & 46.75 & 41.97 & 33.43 & 5.19 & 0.00 \\
& Full-history ICL & 46.48 & 46.12 & 43.18 & 45.24 & 40.38 & 32.09 & 4.64 & -5.77 \\
& VCR & \second{51.38} & \second{51.22} & \second{47.82} & 46.96 & \second{41.98} & \second{33.51} & 3.95 & -1.73 \\
& Hard Reject & 50.24 & 50.43 & 47.06 & 46.75 & 41.63 & 33.13 & 5.92 & +0.56 \\
& Keyword Scrub & 50.23 & 50.40 & 47.03 & \second{47.05} & 41.91 & 33.26 & \best{6.27} & \second{+1.07} \\
& \textbf{\DEC} & \best{53.65} & \best{53.27} & \best{49.74} & \best{48.99} & \best{43.73} & \best{34.85} & \second{5.99} & \best{+3.51} \\
\midrule
\multirow{3}{*}{\shortstack[l]{Gemini\\3-Flash}} & \emph{Profile Adapt.} & \second{51.34} & \second{52.21} & 48.63 & \best{47.60} & \second{42.76} & \second{34.17} & \best{10.53} & \second{0.00} \\
& VCR & 50.72 & \second{52.21} & \second{48.66} & 46.85 & 42.28 & 33.85 & 7.62 & -1.89 \\
& \textbf{\DEC} & \best{53.71} & \best{53.70} & \best{50.03} & \second{47.31} & \best{42.78} & \best{34.19} & \second{9.80} & \best{+1.02} \\
\midrule
\multirow{3}{*}{\shortstack[l]{Qwen3\\4B-Instruct}} & \emph{Profile Adapt.} & 31.34 & 33.42 & 31.22 & \second{31.64} & \second{26.45} & \second{19.99} & 2.89 & 0.00 \\
& VCR & \second{31.45} & \second{33.79} & \second{31.57} & 31.34 & 26.27 & 19.92 & \best{3.26} & \second{+0.37} \\
& \textbf{\DEC} & \best{33.97} & \best{35.42} & \best{33.09} & \best{32.90} & \best{27.65} & \best{20.88} & \second{2.91} & \best{+2.00} \\
\bottomrule
\end{tabular}}
\vspace{5pt}
\includegraphics[width=\linewidth]{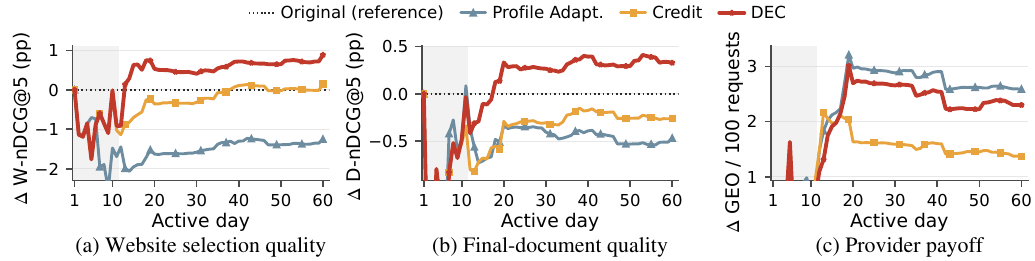}
\captionof{figure}{\textbf{Cumulative outcomes over the $60$-day trajectory}, relative to Original.}
\label{fig:main}
\end{figure}

\subsection{Setup}
\label{sec:setup}

\textbf{Experimental protocol.} We replay \envname's chronological request stream under two protocols: the $30$-day one compares platform-side policies on one fixed set of days, and the $60$-day one follows repeated adaptation to its endpoint over $60$ active days (\Cref{app:equilibrium,app:data-views}). All agent roles use DeepSeek V4-Flash (\Cref{sec:robustness} varies it). We prespecify 10\% of providers as adaptive and the rest as the static web. \emph{Original} fixes all profiles; \emph{Profile Adaptation} lets adaptive agents update them through $F_P$, a history-conditioned continual policy with three actions: keep the profile, add claims supported by past delivery, or add claims without delivery support; one update may combine the latter two (\Cref{app:rewriting-prompt}). \textbf{\DEC} adds the platform-side policy on top.

\textbf{Metrics.} W-nDCG@5 and D-nDCG@5 measure website-selection and final-document quality; Provider GEO measures adaptive providers' position-weighted final-answer exposure per $100$ requests. Let $S_s$ be W-nDCG@5 and $V_s$ be Provider GEO; Original and Profile Adaptation anchor the quality--exposure trade-off per population. Following player welfare in strategic-provider games \citep{benporat2018strategic,madmon2025convergence,xu2026mechanism}, we weight them by net-benefit analysis \citep{vickers2006decision}:
\begin{equation}
\lambda=\max\Big\{0,\ \frac{S_{\mathrm{orig}}-S_{\mathrm{adapt}}}{V_{\mathrm{adapt}}-V_{\mathrm{orig}}}\Big\},
\qquad
G(s)=(S_s-S_{\mathrm{adapt}})+\lambda(V_s-V_{\mathrm{adapt}}).
\label{eq:joint-gain}
\end{equation}
$\lambda$ converts provider exposure into W-nDCG@5 units, so Welfare Gain $G(s)$ is the net benefit above that trade-off line (\Cref{sec:game}), with Profile Adaptation as the zero reference (\Cref{app:metrics,app:weights}). Each backbone and protocol has its own $\lambda$, so Welfare Gain is compared within one. Paired $95\%$ confidence intervals are reported in \Cref{app:full60,app:thirtyday}, with the resampling procedure in \Cref{app:uncertainty}.

\textbf{Baselines.} \emph{Profile Adaptation} is our main baseline; we also compare \DEC with four platform-side ones: \textbf{Full-history ICL} places past service records in the selection context \citep{asawa2026clbench}; \textbf{VCR} uses verifiable-content reward and suspicion signals from successive profile versions \citep{xu2026mechanism}; \textbf{Keyword Scrub} removes fixed GEO-style phrases; \textbf{Hard Reject} filters by the current profile's suspicion score. Each follows its source configuration and shares $F_P$ (\Cref{app:user-baselines,app:repro}). Additional history-based heuristic controls are reported in \Cref{app:user-baselines,app:beta-diagnostic}.

\subsection{Main results}
\label{sec:main}
\label{sec:comparisons}
\label{sec:alternatives}
\label{sec:platform-baselines}
\label{sec:longrun}

\textbf{Provider-side strategies.} Profile Adaptation reaches the harmful endpoint predicted in \Cref{sec:game}. Among five provider strategies run on the same $30$-day population without platform-side intervention, it is the only one with a material gain and the only one that costs the user side (\Cref{tab:provider-baselines,app:provider-baselines}); the harm therefore comes from the best-performing evaluated strategy.

\textbf{\DEC improves both sides.} In \Cref{tab:platform}, Full-history ICL, Keyword Scrub, and Hard Reject do not recover the user-side loss, and VCR recovers limited quality at a cost in provider exposure; Keyword Scrub's phrases miss adaptive profiles and weaken static ones. \DEC alone improves website-selection quality while preserving the exposure gain, with the highest Welfare Gain. \DEC also improves generated-answer correctness, completeness and citation groundedness while retaining most citation-based provider exposure (\Cref{app:generated-answers}).

\textbf{The pattern holds over the long-horizon trajectory.} Over active days $21$--$60$ of the $60$-day stream, \DEC Pareto-dominates Original in \Cref{fig:main} and ends at a Welfare Gain of $+2.01$ points (\Cref{app:full60,app:weights}). Market openness is preserved: the Top-10 share is the same under \DEC and Profile Adaptation, and less concentrated than under Original (\Cref{app:market-structure}).

\begin{table}[!t]
\centering
\caption{\textbf{Provider-side strategies.} Bold: best under the provider's objective.}
\label{tab:provider-baselines}
\footnotesize
\setlength{\tabcolsep}{2.2pt}
\renewcommand{\arraystretch}{0.92}
\setlength{\abovecaptionskip}{4pt}
\setlength{\belowcaptionskip}{2pt}
\begin{tabular}{lrrrr}
\toprule
& \multicolumn{2}{c}{Provider exposure} & \multicolumn{2}{c}{User-side quality} \\
\cmidrule(lr){2-3}\cmidrule(lr){4-5}
Provider-side strategy & At selection & Provider GEO & W-nDCG@5 & D-nDCG@5 \\
\midrule
\emph{Original} (static) & 16.27 & 4.17 & 52.61 & 42.11 \\
GEO Keyword Stuffing \citep{aggarwal2024geo} & 16.34 & 4.23 & 52.94 & 42.21 \\
GEO Fluency Optimization \citep{aggarwal2024geo} & 16.02 & 4.22 & 53.13 & 42.99 \\
MPMA Authority \citep{wang2026mpma} & 13.03 & 3.24 & 52.79 & 42.41 \\
FedRAG Universal \citep{mu2026routinghijack} & 10.11 & 2.76 & 53.01 & 42.16 \\
\textbf{Profile Adapt.} & \best{22.07} & \best{5.19} & 51.01 & 41.97 \\
\bottomrule
\end{tabular}
\end{table}

\begin{figure}[!t]
\centering
\includegraphics[width=\linewidth]{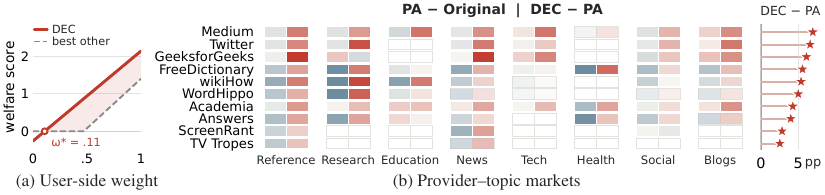}
\caption{\textbf{Robustness within a run.} (a) Welfare score vs.\ user-side weight; \DEC leads above $\omega^*$. (b) W-nDCG@5 change per provider--topic market, paired within the market; PA denotes Profile Adaptation, blue a loss and red a gain; the last column is each provider across all topics.}
\label{fig:robustness}
\end{figure}

\subsection{Ablation}
\label{sec:ablation}

\textbf{Credit and Withhold are complementary.} Credit enriches the selection context with verified delivery evidence related to the current request, helping the selector identify websites that have served similar needs. On this trajectory, Credit alone raises W-nDCG@5 by $1.40$ points over Profile Adaptation. Withhold alone yields smaller gains of $0.21$ W-nDCG@5 and $0.27$ D-nDCG@5 points: it can remove a repeatedly non-delivering website, but cannot supply the positive evidence needed to backfill the freed opportunity with a better one. Combining the two closes this gap: Withhold frees the opportunity, and Credit supplies verified evidence for its replacement. Adding Withhold to the selection Credit has shaped contributes another $0.74$ W-nDCG@5 points; full \DEC is higher than either single-action variant in W-nDCG@5, D-nDCG@5, Provider GEO, and Welfare Gain (\Cref{tab:ablation}). Paired comparisons are in \Cref{app:full60,app:funnel}.

\begin{table}[!t]
\centering
\caption{\textbf{Credit/Withhold ablation over $60$ days.} Bold marks the best continual-rewriting variant.}
\label{tab:ablation}
\footnotesize
\setlength{\tabcolsep}{3.4pt}
\renewcommand{\arraystretch}{0.92}
\setlength{\abovecaptionskip}{4pt}
\setlength{\belowcaptionskip}{2pt}
\begin{tabular}{lrrrr}
\toprule
Variant & W-nDCG@5 & D-nDCG@5 & Provider GEO & Welfare Gain \\
\midrule
\emph{Original} & 51.91 & 41.32 & 5.11 & 0.00 \\
Profile Adapt. & 50.66 & 40.85 & \best{7.60} & 0.00 \\
\quad + Withhold & 50.87 & 41.12 & 7.14 & $-0.02$ \\
\quad + Credit & 52.06 & 41.20 & 6.36 & $+0.78$ \\
\quad \textbf{+ Credit and Withhold (\DEC)} & \best{52.80} & \best{41.56} & 7.34 & \best{+2.01} \\
\bottomrule
\end{tabular}
\end{table}

\begin{table}[!t]
\centering
\caption{\textbf{Full-trajectory robustness.} Welfare Gain uses the shared Original trajectory and the setting-specific Profile Adaptation trajectory as anchors.}
\label{tab:trajectory-robustness}
\footnotesize
\setlength{\tabcolsep}{4.4pt}
\renewcommand{\arraystretch}{1.08}
\setlength{\abovecaptionskip}{4pt}
\setlength{\belowcaptionskip}{2pt}
\begin{tabular}{@{}llrrrr@{}}
\toprule
Setting & System & W-nDCG@5 & D-nDCG@5 & Provider GEO & Welfare Gain \\
\midrule
\multirow{2}{*}{Less-restricted updater} & \emph{Profile Adapt.} & 49.68 & 38.98 & 6.89 & 0.00 \\
& \textbf{\DEC} & \best{51.21} & \best{40.45} & \best{7.09} & \best{+1.78} \\
\midrule
\multirow{2}{*}{Independent replay 1} & \emph{Profile Adapt.} & 51.51 & 40.75 & \best{7.39} & 0.00 \\
& \textbf{\DEC} & \best{52.89} & \best{42.06} & 7.01 & \best{+1.31} \\
\midrule
\multirow{2}{*}{Independent replay 2} & \emph{Profile Adapt.} & 50.76 & 40.56 & \best{7.89} & 0.00 \\
& \textbf{\DEC} & \best{52.76} & \best{41.82} & 6.87 & \best{+1.58} \\
\bottomrule
\end{tabular}
\end{table}

\subsection{Robustness}
\label{sec:robustness}

Across alternative system configurations and execution conditions, \DEC consistently improves user-side quality while preserving provider benefit. The result also holds across the markets induced by the resulting trajectories.

\textbf{Backbone robustness.} Across all three backbones, \DEC improves website-selection quality over Profile Adaptation and attains the highest Welfare Gain among the compared systems, while preserving most of rewriting's exposure gain (\Cref{tab:platform}); counting exposure of all websites also ranks it first (\Cref{app:metrics}). This improvement is therefore not tied to one model family (\Cref{app:thirtyday}).

\textbf{Updater robustness.} The result persists with a less restricted profile updater: \DEC improves both user-side quality measures over Profile Adaptation, reduces endpoint unbacked claims by $49\%$, and preserves long-horizon profile stability (\Cref{tab:trajectory-robustness,app:updater-robustness}).

\textbf{Changing the adaptive websites.} On a second group of adaptive websites, \DEC again improves both user-side quality measures over Profile Adaptation while retaining most of its exposure gain, for a Welfare Gain of $+1.90$. The same two-sided benefit holds when profile adaptation becomes more widespread and the adaptive share rises from $10\%$ to $30\%$ (\Cref{app:alternative-adaptive-websites}).

\textbf{Across-run robustness.} We add two complete $60$-day replays conducted in different periods. \DEC improves both user-side quality measures over Profile Adaptation in each replay, while between-run variation in provider exposure contracts from $21.7\%$ in the first ten days to $3.0\%$ in the final ten days. Welfare Gain remains positive in all three complete trajectory checks (\Cref{tab:trajectory-robustness,app:across-run}).

\textbf{Additional checks and best response.} The result is stable across welfare weights, markets, topics and request difficulty: \DEC leads whenever user-side quality receives at least $32.9\%$ of the weight (\Cref{fig:robustness}a, \Cref{app:weights}); it improves website-selection quality in nearly every provider--topic market and for every adaptive provider (\Cref{fig:robustness}b); and the gain holds across topics and request difficulty (\Cref{app:topic-heterogeneity,app:additional-robustness}). All provider strategies so far optimize profile content; we also test malicious prompt injection \citep{qian2025rankingblindspot}. Without intervention it is the best response among these strategies and the most damaging. Intervention makes it unprofitable; $F_P$ is again the best response among evaluated strategies, yielding a healthier outcome (\Cref{app:provider-baselines}).

\begin{figure}[!t]
\centering
\includegraphics[width=\linewidth]{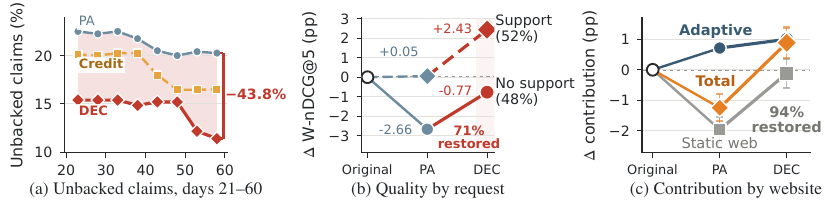}
\caption{\textbf{Profile composition and website matching over the $60$-day trajectory.}}
\label{fig:capability-contribution}
\refstepcounter{subfigure}\label{fig:unbacked-claims}%
\refstepcounter{subfigure}\label{fig:quality-by-supply}%
\refstepcounter{subfigure}\label{fig:source-contributions}%
\end{figure}

\subsection{Why \DEC works}
\label{sec:healthier-ecosystem}
\label{sec:why}

Both \DEC actions reallocate opportunities from recorded delivery. We examine why this turns the harmful endpoint toward a healthier outcome by reducing information asymmetry on both sides: content agents write more delivery-backed claims, and the user agent selects better.

\subsubsection{Delivery-grounded profile rewriting}
\label{sec:better-profiles}

\textbf{Rewriting toward delivery-backed claims.} The experiments match the analysis in \Cref{sec:game}. Under Profile Adaptation, profiles reach a harmful stable point under repeated $F_P$ updates and retain many unbacked claims (\Cref{app:profiles}): websites claim capabilities without corresponding delivery, and the endpoint empirically meets our approximate rewriting-equilibrium criterion (\Cref{app:equilibrium}). Content agents still seek exposure, but \DEC changes how evidence accumulates for the two kinds of claims: Credit gives topics with cited documents more later opportunities, while Withhold stops repeatedly non-delivering topics from occupying them, so delivered topics gain a growing evidence advantage. Limited profile capacity turns this advantage into claim choices in the next rewrite, cutting the endpoint share of unbacked claims by $43.8\%$ relative to Profile Adaptation; after the unstable first third of the run (\Cref{app:equilibrium}), \DEC stays below Profile Adaptation (\Cref{fig:unbacked-claims}).

\textbf{Two-sided benefits.} Fewer unbacked claims translate into better later website selection. We hold the selector fixed and give it only the provider-written profiles from each policy, with no evidence cards: those from \DEC raise W-nDCG@5 by $2.47$ points over those from Profile Adaptation in the same round. For content agents, \DEC raises final-use conversion and cuts the share of adaptive websites' final-answer exposure that falls on irrelevant matches, while retaining $89.7\%$ of Profile Adaptation's exposure gain over Original (\Cref{app:funnel}). Better user-side selection and preserved provider-side benefit together yield a healthier ecosystem outcome.

\begin{figure}[!t]
\centering
\includegraphics[width=\linewidth]{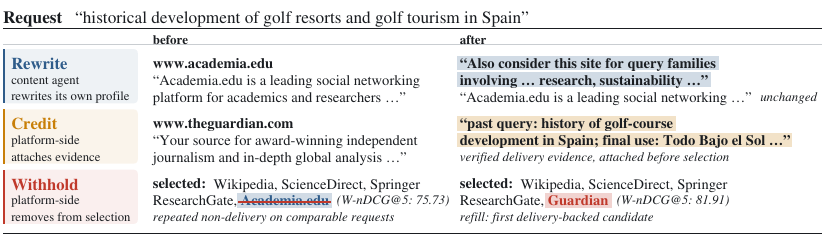}
\caption{\textbf{One observed request}, with the content agent's Rewrite and \DEC's Credit and Withhold.}
\label{fig:where}
\end{figure}

\subsubsection{Better website matching}
\label{sec:better-matching}

\textbf{Narrowing the claims--delivery gap.} Under Profile Adaptation, self-written claims cannot be verified in advance, so websites that appear relevant but cannot deliver effectively take up limited opportunities and displace sources that can answer the request. \DEC uses market-wide history so the user agent can consider what a website has delivered as well as what it claims: Credit adds delivery evidence, while Withhold reduces repeated-non-delivery matches. History alone is insufficient: with identical candidates and query-conditioned history, summarizing it into a score with a standard reputation rule \citep{josang2002beta} and reranking lowers W-nDCG@5 by a further $2.24$ points on Profile Adaptation (\Cref{app:beta-diagnostic}). Together Credit and Withhold turn this history into evidence and narrow the gap, so opportunities reach websites more likely to answer.

\textbf{Recovering displaced static websites.} Profile Adaptation's harm concentrates on requests with no relevant adaptive website (\Cref{fig:quality-by-supply}): quality falls below Original entirely through lost static contribution (\Cref{fig:source-contributions,tab:supply-contributions}), and \DEC repairs most of it. It returns only a small share of the displaced static opportunities yet recovers nearly all of the lost contribution, so the returned opportunities reach the websites that contribute most. Market participation stays as broad as under Profile Adaptation, and static-web exposure is nearly unchanged (\Cref{app:market-structure}).

\textbf{Improving useful adaptive contributions.} On requests with a relevant adaptive website, Profile Adaptation's gain in adaptive contribution is almost entirely offset by its static loss, leaving quality near Original. \DEC gives the user agent more reliable capability signals, raising both contributions and lifting W-nDCG@5 above Original (\Cref{fig:quality-by-supply}). \DEC therefore strengthens useful adaptive websites and restores static contribution, rather than suppressing them uniformly.

\textbf{Creating a positive feedback loop.} \DEC closes the two histories into one loop: provider-local history helps content agents write profiles that better reflect actual delivery, while market-wide history gives the user agent evidence beyond self-written claims. Better selections then produce service experience that later rewriting and selection can use, and over the $60$ active days this loop sustains the gains in request--website matching (\Cref{app:market-structure}) and welfare (\Cref{app:weights}).

\subsection{Case study}
\label{sec:heterogeneous-markets}

The Rewrite, Credit, and Withhold examples in \Cref{fig:where} illustrate the profile-driven selection mismatch exposed by Profile Adaptation and how \DEC responds to it using recorded delivery outcomes. For this request about the history of golf tourism in Spain, Academia.edu wins a scarce website-selection opportunity with its rewritten profile, but its repeated selection for comparable requests has not converted into final use. \DEC withdraws this opportunity and backfills it with the Guardian, which has a recorded final-use delivery, improving W-nDCG@5 for this request. \Cref{app:mechanism-cases} gives separate request-level traces for the three behaviors.

\subsection{Discussion and future work}
\label{sec:discussion}

Our work focuses on an understudied stage of the Agentic Web: how content agents rewrite profiles to compete for website selection before retrieval. The main experiments therefore hold documents fixed to study this profile-level competition over a long horizon. In realistic deployments, content agents may also improve or manipulate documents together with their profiles.

\begin{table}[!t]
\centering
\caption{\textbf{Joint profile--document optimization under the $30$-day protocol.}}
\label{tab:joint-optimization}
\footnotesize
\setlength{\tabcolsep}{2.7pt}
\begin{tabular}{@{}lrrrrrrr@{}}
\toprule
System & W@3 & W@5 & W@10 & D@3 & D@5 & D@10 & Provider GEO \\
\midrule
\emph{PA-Joint} & 50.70 & 50.94 & 47.56 & 45.63 & 40.67 & 32.24 & \best{6.94} \\
VCR-Joint & \second{51.72} & \second{51.42} & \second{48.00} & \second{46.21} & \second{41.07} & \second{32.82} & 4.66 \\
\textbf{\DEC-Joint} & \best{52.24} & \best{51.97} & \best{48.51} & \best{47.12} & \best{41.84} & \best{33.34} & \second{5.42} \\
\bottomrule
\end{tabular}
\end{table}

We test a broader provider action space, in which each adaptive content agent revises its profile and multiple documents previously retrieved from its own corpus and returned in past interactions. PA-Joint applies this joint action; VCR-Joint and \DEC-Joint extend VCR and \DEC to both profile and document adaptation under the same action space and update budget (\Cref{app:joint-optimization}). PA-Joint raises Provider GEO by $33.7\%$ over profile-only adaptation. \DEC-Joint leads all user-side measures and outperforms VCR-Joint on both user-side quality and Provider GEO (\Cref{tab:joint-optimization}), showing that \DEC remains effective when content agents adapt both profiles and documents. Future work can extend joint adaptation to adding useful documents for emerging user needs.

\FloatBarrier

\section{Conclusion}
\label{sec:conclusion}

We study agent ecosystems in the Agentic Web, where content agents compete for website selection by rewriting the profiles a user agent reads. We build \envname, a simulation framework and benchmark. Provider adaptation stabilizes with higher exposure but lower website-selection quality. We therefore design \DEC, a platform-side policy that uses recorded delivery outcomes to reallocate later opportunities, restoring quality above the static baseline while retaining most exposure gains. More broadly, website selection allocates both current opportunities and future learning evidence, making it a central mechanism-design lever for healthier agent ecosystems.

\bibliography{references}
\bibliographystyle{references}

\clearpage
\appendix
\FloatBarrier
\section{Evaluation metrics}
\label{app:metrics}

The appendix follows the evidence chain: Section A defines the evaluation metrics; Sections B and C specify the benchmark, the dynamic game, and the method and baseline details; Sections D--F report long-horizon outcomes, mechanism and ecosystem evidence, and robustness; and Sections G--J give the proofs, prompts, reproducibility details, and scope.

For request $q$ and an ordered list $\mathbf x=(x_1,\ldots,x_k)$ with frozen evaluator gains $g_q$, we use linear-gain nDCG:
\begin{equation}
\operatorname{nDCG@}k(q,\mathbf x)=
\frac{\sum_{i=1}^{k}g_q(x_i)/\log_2(1+i)}
{\operatorname{IDCG@}k(q)}.
\label{eq:ndcg}
\end{equation}
Here $\operatorname{IDCG@}k(q)$ is the maximum discounted gain for $q$ at cutoff $k$. W-nDCG applies this score to the selector-ranked websites and their website-level gains; D-nDCG applies it to the documents cited in the final answer and their document-level grades. System values are request-level macro averages reported on a $0$--$100$ scale.

For an ordered list of length $n\ge1$, the position discount is normalized over that list:
\begin{equation}
\nu_i=\frac{1/\log_2(1+i)}{\sum_{j=1}^{n}1/\log_2(1+j)},
\qquad \sum_{i=1}^{n}\nu_i=1,
\label{eq:position}
\end{equation}
\Cref{eq:exposure} applies $\nu_i$ to each position of the ranked list.

For system $s$ on request population $Q$, let $\mathcal A$ be the ten adaptive websites and $E_t^s(w,q)$ their position-weighted exposure in that run. Provider GEO per $100$ requests is
\begin{equation}
V_s=\frac{100}{|Q|}\sum_t\sum_{q\in Q_t}\sum_{w\in\mathcal A}E_t^s(w,q).
\label{eq:provider-geo}
\end{equation}
$V_s$ covers the ten adaptive content agents, which form the provider side of the game in \Cref{sec:game}; static content agents keep fixed profiles and take no action in it. This follows welfare definitions over the players of games with strategic content providers and of LLM ad auctions, which sum the payoffs of the strategic participants and measure users' relevance or output quality separately \citep{benporat2018strategic,madmon2025convergence,madmon2025stability,hajiaghayi2024ad,xu2026mechanism}. Summing exposure over all one hundred websites would add an almost constant term: the impressions of all citations in an answer sum to one \citep{aggarwal2024geo}, so cumulative exposure is nearly constant across systems \citep{do2021twosided}, and a welfare over every website orders the systems almost entirely by user-side quality, on which \DEC ranks first in every platform-side comparison. The static web's contribution to W-nDCG@5 (\Cref{tab:source-contributions}) is counted in user-side quality; its exposure is reported as an allocation outcome, which \DEC leaves nearly unchanged relative to Profile Adaptation while returning static-web seats (\Cref{app:market-structure}).

\paragraph{Welfare Gain and its calibration.} Write $S_s$ for W-nDCG@5 and retain the subscripts $\mathrm{orig}$ and $\mathrm{adapt}$ of \Cref{eq:joint-gain}. The corresponding empirical trade-off line is $B(V)=S_{\mathrm{adapt}}-\lambda(V-V_{\mathrm{adapt}})$, so $G(s)=S_s-B(V_s)$ is the W-nDCG distance above that line, and $G(s)>0$ places a system above this observed trade-off. $\lambda$ is the W-nDCG@5 given up per unit of Provider GEO between the two anchors, the opportunity cost of exposure. When rewriting lowers W-nDCG@5, the line passes through both anchors and $G(\mathrm{orig})=G(\mathrm{adapt})=0$; the two anchors fix the reference line, so these zeros are a normalization. When it does not, exposure costs no user-side quality and $\lambda=0$, as the exchange rate of net-benefit analysis is non-negative \citep{vickers2006decision}; $G$ is thus non-decreasing in both sides in every population. Estimating it from the observed exchange between Original and Profile Adaptation parallels estimating cost-effectiveness thresholds from empirical marginal expenditure--outcome relationships \citep{claxton2015threshold}. Calibrating it this way avoids fixing a weight by assumption, which would make the ordering of systems a property of that choice; two-sided ranking work makes the same point when it replaces a single welfare function with a dominance criterion \citep{do2021twosided}. This net-benefit construction \citep{vickers2006decision} places website-selection quality and provider exposure on a common scale. $\lambda$ thus serves as a reference rate for comparing systems within one population.

\paragraph{A weight-free comparison.} \DEC's welfare improvement holds without a weight relative to Original on the $60$-day trajectory, under a range of non-negative weights in every setting, and with exposure summed over all one hundred websites (above). A system that improves both sides relative to Original ranks above it under every monotone welfare function. On the $60$-day trajectory, \DEC improves both website-selection quality and Provider GEO, while Profile Adaptation gains exposure but loses quality (\Cref{tab:long-horizon-paired}). This comparison uses neither a weight nor the calibrated line. Across all four settings, \DEC ranks first whenever user-side quality receives at least $32.9\%$ of the weight; equal weighting assigns $50\%$ and meets this condition (\Cref{app:weights}).

\paragraph{Comparability and uncertainty.} Because $\lambda$ is fixed by the two anchors of one population, Welfare Gain is compared only within that population. Profile rewriting raises provider exposure on all three backbones; on Qwen it does not lower website-selection quality, so $\lambda=0$ there and Welfare Gain is the W-nDCG@5 difference from Profile Adaptation. The ordering under positive weights on exposure is in \Cref{app:weights}; on Qwen it ranks \DEC first for every $\omega\ge0.178$. The body reports point estimates; the $95\%$ confidence intervals are in \Cref{app:full60,app:funnel,app:thirtyday} and the resampling procedure in \Cref{app:uncertainty}.

\paragraph{Generated-answer evaluation.}\label{app:generated-answers}
Following the AutoGEO protocol \citep{wu2025autogeo}, we generate natural-language answers with citations for all $4{,}675$ requests in the $60$-day evaluation of Profile Adaptation and \DEC. We use a G-Eval rubric \citep{liu2023geval}, with GPT-4o-mini as the judge, to evaluate correctness, completeness and citation groundedness.

\begin{table}[!h]
\centering
\caption{\textbf{Generated-answer quality.}}
\label{tab:generated-answers}
\footnotesize
\setlength{\tabcolsep}{3.0pt}
\begin{tabular}{@{}lrrrr@{}}
\toprule
System & Correct. & Complete. & Grounded. & Provider GEO \\
\midrule
Profile Adapt. & 4.554 & 4.548 & 4.529 & \best{6.18} \\
\textbf{\DEC} & \best{4.571} & \best{4.566} & \best{4.548} & 5.83 \\
\bottomrule
\end{tabular}
\end{table}

Citation groundedness judges whether answer claims are supported by their assigned citations, while Provider GEO uses the citations assigned in generated answers. \DEC is higher on all three answer measures, and the 95\% paired bootstrap CIs are $[0.0095, 0.0255]$, $[0.0114, 0.0232]$, $[0.0107, 0.0269]$. Across paired requests, a $0.1$-point D-nDCG@5 increase is associated with $+0.119$ correctness (95\% CI $[0.098,0.140]$), $+0.120$ completeness ($[0.099,0.142]$), and $+0.122$ citation groundedness ($[0.101,0.143]$). Citations in the generated answers retain $94.4\%$ of Profile Adaptation's Provider GEO. Together, these results show that \DEC improves generated-answer quality while retaining most provider exposure.

\FloatBarrier
\section{Benchmark construction and game specification}
\label{app:construction}

\subsection{Request stream and evaluation views}
\label{app:data-views}

The local DeepResearchGym log first passes the format checks, internal-test filtering, and anomalous-duplicate handling of the source data protocol \citep{ning2026agenticsearch}. It is then split into sessions by time and semantic continuity. After this source-log cleaning, there are $556$ HMAC-IP client proxies and $14{,}369{,}746$ query events; one query event is one \texttt{/search} API call.

We then use two items of client-level metadata to construct a longitudinal request stream. Each client must be active on at least $3$ days, so its trajectory spans several end-of-round updates; this follows the use of activity span to retain complete trajectories in longitudinal query-log studies \citep{teevan2007repeat}. A client must also average at most $50$ semantic sessions per active day, which removes concentrated high-throughput traffic. Search intensity and repetition help distinguish automated from ordinary use patterns \citep{buehrer2008automated}, while daily intensities in natural-use logs provide an order-of-magnitude reference for this dataset-specific boundary \citep{maheux2026chatbots}. The filters retain $87$ clients and all their events, yielding $9{,}771$ requests, $3{,}726$ sessions, and $102$ active days.

\paragraph{Evaluation views.}

\begin{table}[t]
\centering
\caption{\textbf{Chronological evaluation views derived from the source request stream.}}
\label{tab:evaluation-views}
\small
\setlength{\tabcolsep}{2pt}
\begin{tabular}{>{\raggedright\arraybackslash}p{0.23\linewidth}rr>{\raggedright\arraybackslash}p{0.43\linewidth}}
\toprule
View & Requests & Active rounds & Membership rule \\
\midrule
Filtered complete request stream & 9{,}771 & 102 & all valid events from the $87$ clients \\
Evaluable request set & 9{,}043 & 102 & at least one frozen-pool document has grade above $0$ \\
60-day common evaluation & 4{,}675 & 60 & every evaluable request in the first $60$ active days \\
Comparison view, $\le100$ per round & 4{,}024 & 102 & up to $100$ requests per round in event order \\
30-day comparison prefix & 1{,}465 & 30 & first $30$ non-empty active days of that view \\
\bottomrule
\end{tabular}
\end{table}

We construct two evaluation views with different purposes from the same chronological request stream. The $30$-day comparison uses the first $30$ non-empty active days of the comparison view with at most $100$ requests per round, giving $1{,}465$ requests and $626$ sessions. The $60$-day analysis uses every evaluable request in the first $60$ active days, giving $4{,}675$ requests and $1{,}332$ sessions. Each comparison pairs requests only across systems that share a view, and requests are not removed based on system completion.

\begin{figure}[t]
\centering
\includegraphics[width=\linewidth]{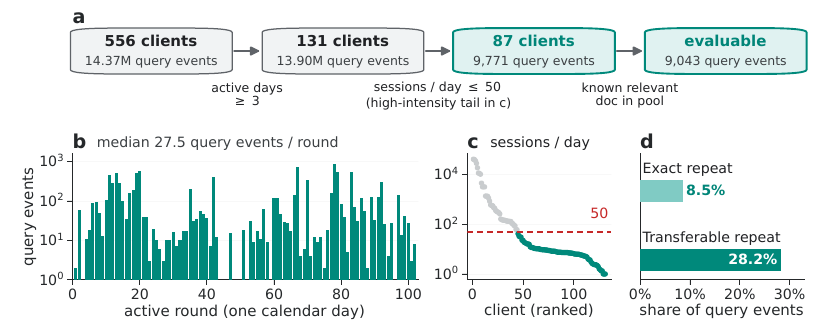}
\caption{\textbf{Construction of \envname and its request stream.} (a) Data funnel; (b) daily volume; (c) client activity and threshold; (d) historical repetition and transferable delivery.}
\label{fig:env}
\end{figure}

One UTC calendar day is one round, and \Cref{fig:env} shows the data funnel and the resulting stream. Repeated demand is what lets recorded delivery inform later requests: $8.53\%$ of requests repeat an exact request from an earlier active day, and $28.22\%$ have a semantically comparable earlier request on which some website delivered. Both shares were measured after the timeline was frozen and played no part in inclusion.

\subsection{Relevance judgements}
\label{app:judgements}

\paragraph{Candidate construction and website gain.} For each distinct request, BM25 and MiniCPM-dense each retrieve their top $100$ documents, whose union contains $185.5$ documents on average. A BGE reranker ranks the complete union, and GPT-4o grades its top $20$ query--document pairs independently on the UMBRELA $0$--$3$ scale. Thus, the candidate-recall range is this two-retriever union, and top $20$ is the uniform fixed judgement depth for all requests. This follows benchmarks that combine multi-route candidate generation with fixed shallow judgement: MIRACL assesses fused retrieval at depth $10$, while Utility-Focused LLM Annotation and FreshStack use multi-system pools capped at $20$ for LLM judgement \citep{zhang2023miracl,zhang2025utility,thakur2025freshstack}. Retrieval studies whose subject is not relevance assessment also label their own evaluation data with UMBRELA: BAGEL generates TravelDest query--passage labels with a binary UMBRELA prompt and evaluates retrieval on them \citep{kim2026bagel}, and a study of positional bias in long-document ranking estimates where relevant passages occur with UMBRELA relevance judges \citep{boytsov2025positional}.

Our pipeline judges $150{,}140$ query--document pairs, and the resulting frozen qrels map to $9{,}043$ evaluation events. For request $q$ and website $w$, $g$ takes the original UMBRELA grade $0$, $1$, $2$, or $3$. Let $g_{(1)}\ge g_{(2)}\ge\cdots$ be the positive grades of all judged documents from that website. Its gain is $g^W(q,w)=\sum_j g_{(j)}/\log_2(j+1)$. D-nDCG likewise uses the original document grade.

All systems share these frozen qrels. They define the evaluation population and are used to compute metrics after each run, but do not enter any system trajectory. This separation reflects the deployment setting: per-website or per-document relevance and satisfaction labels are not produced after every request, clicks are affected by display position, and feedback on a generated answer is hard to attribute to one website \citep{joachims2017unbiased,dai2025nextsearch}. During a trajectory, website selection and profile updates can use the request, website profiles, and observed delivery records, such as which returned documents enter the final answer. The qrel grades and all metrics derived from them remain visible only to the evaluator.

\paragraph{System-level agreement with existing labels.} Prior evaluations validate automatic relevance labels through system-ranking agreement with human labels and blinded in-domain audits \citep{chen2025airbench,upadhyay2025largescale,yin2025disastir,daiber2025dispatchqa}. We use these two complementary checks: system-level agreement with existing labels and an in-domain blinded agreement audit.

To test whether GPT-4o labels preserve system conclusions supported by existing labels, we use episodes in the \envname log that can be linked back to existing labels from MS MARCO Web Search, hereafter MARCO, and Researchy. MARCO provides query--document click labels for web search, while Researchy provides complex, multi-perspective questions and clicked-document labels \citep{chen2024msmarcoweb,rosset2024researchy}. The two datasets therefore give these requests an external reference independent of GPT-4o. These episodes validate the relevance judge and are not the main request stream of the long-horizon experiment. The external test covers $160$ episodes, $155$ distinct requests, and $3{,}478$ query--document pairs. For each request, we take $\mathrm{BM25@100}\cup\mathrm{dense@100}$, rerank it with BGE, and judge $\mathrm{BM25@10}\cup\mathrm{dense@10}\cup\mathrm{BGE@10}$, which averages $22.4$ documents. Existing and GPT-4o labels separately evaluate three frozen retrieval systems: BM25, dense, and BGE. Under the GPT-4o labels both metrics read the original UMBRELA grades: D-nDCG@10 uses document gains and W-nDCG@5 uses discounted website gain. The $80$ episodes at q0--39 form the development block, and the $80$ at q40--79 form the confirmation block. Differences are paired by episode within each dataset--block and bootstrapped $10{,}000$ times with fixed seed 20260729.

Under the existing labels, $16$ system-pair--metric comparisons have $95\%$ confidence intervals that exclude zero; none changes direction under the GPT-4o labels. There are four such comparisons in each dataset--block combination, all comparing BM25 with dense or BGE across the two metrics. Researchy retains the complete BGE $>$ dense $>$ BM25 ordering in both blocks, with Kendall correlation $1.0$, and every clear semantic-versus-BM25 comparison on MARCO also retains its direction. Thus, the GPT-4o qrels reproduce the retrieval-system conclusions supported by the existing labels in both the development and confirmation blocks.

\paragraph{In-domain blinded agreement audit.} The second check follows the in-domain stage of the DispatchQA validation design: after the external test, an independent judge re-grades one in-domain sample under blind conditions, and agreement is reported between that judge and the production judge \citep{daiber2025dispatchqa}. We randomly sample $24$ pairs at each GPT-4o grade from the final Wild candidate pool, for $96$ total, following the balanced graded-audit design used in prior relevance-judgement work \citep{faggioli2023perspectives}. The second judge is a human annotator applying the same UMBRELA $0$--$3$ rubric, who sees only the query and the full document text and is blinded to the GPT-4o grade, candidate source, website metadata, retrieval rank, and the production labels. The sample is drawn and frozen before the audit, and the two label sets are joined only after all $96$ pairs have been graded. We report exact agreement and unweighted Cohen $\kappa$ on the four-level labels, then Cohen $\kappa$ again with grades $0$--$1$ and $2$--$3$ grouped as non-answer-bearing and answer-bearing. This binary split is only an agreement diagnostic; D-nDCG and W-nDCG continue to use the original $0$--$3$ grades directly.

GPT-4o and the human judge agree exactly on $56/96$ ($58.3\%$) of the four-level labels, with Cohen $\kappa=0.44$. At the answer-bearing boundary, Cohen $\kappa=0.73$. Of the $40$ four-level disagreements, $39$ differ by one grade and one by two grades; no pair receives grades $0$ and $3$ from the two judges. The residual variation therefore lies almost entirely in adjacent-grade calibration, while agreement is stronger at the answer-bearing boundary.

\paragraph{Robustness to unjudged documents.} We test whether documents outside the top-$20$ judgement pool change the system comparison, a known issue in retrieval evaluation \citep{froebe2023bootstrapped}. We recompute both metrics on condensed lists, which drop every unjudged entry before scoring \citep{sakai2007bpref}. On the $30$-day protocol, \DEC over Profile Adaptation is $+1.71$ points of W-nDCG@5 (95\% CI $[+0.75,+2.71]$) and $+1.41$ points of D-nDCG@5 ($[+0.47,+2.36]$), both intervals excluding zero.

\subsection{Full dynamic-game specification}
\label{app:game-specification}

\Cref{sec:game} retains the players, observations, interaction, and objectives needed to understand the simulation. Here we give the complete state transition and utility definitions. At the start of round $t$, the user agent selects from current request $q$ and registry $R_t$ before returned documents are available. The profile claims it reads are therefore unverifiable signals sent before its action, as in cheap-talk games \citep{crawford1982strategic}. Below, $\pi_U$ denotes the platform-side policy under study; its input signature includes the platform-maintained market-wide history $M_{t-1}^U$, which $\pi_U^{\mathrm{base}}$ ignores. $M_{t-1}^U$ contains all service records through round $t-1$. For website $w$, $M_{t-1}^P(w)$ is the ordered provider-local service history containing only its selected requests through round $t-1$, with event $(q,\mathcal D(q,w),\delta(q,w))$ for request $q$; write $\rho_{1:t}(w)=(\rho_1(w),\ldots,\rho_t(w))$ for its deployed-profile trajectory.

The user agent selects $L_t(q)$ through $\pi_U$, while each adaptive content agent submits its next profile through $F_P$ at round end. After all selections in round $t$ execute, $Y_t$ is the platform-maintained market-wide service-record sequence. For each website $w$, $Y_t(w)$ is its ordered provider-visible record sequence and contains only requests sent to $w$ and $w$'s own return and citation outcomes. Selection thus determines which outcomes enter later histories, the feedback loop studied in recommender systems \citep{chaney2018confounding}. Let $\operatorname{Append}(M,Y)$ append record sequence $Y$ to history $M$ in chronological order. The coupled recursion is
\begin{equation}
\left\{
\begin{aligned}
L_t(q)&=\pi_U(q;R_t,M_{t-1}^U),\\
M^U_t&=\operatorname{Append}(M^U_{t-1},Y_t),\\
M^P_t(w)&=\operatorname{Append}(M^P_{t-1}(w),Y_t(w)),\\
\widehat\rho_{t+1}(w)&=F_P\!\left(\rho_{1:t}(w),M^P_t(w)\right).
\end{aligned}
\right.
\label{eq:memory}
\end{equation}
Static websites use $\widehat\rho_{t+1}(w)=\rho_t(w)$; all $\widehat\rho_{t+1}(w)$ are next-round submissions. These submissions form the selector-visible registry $R_{t+1}$.

Let $\mathbf x_t(q)=(x_1,\ldots,x_k)$ denote the ordered outcomes the user agent produces for request $q$, $\rel(q,x_i)$ their graded relevance, and $\nu_i$ the normalized logarithmic position weight defined in \Cref{eq:position}. At the final-answer stage, let $(d_1,\ldots,d_n)$ be the cited documents in order. Website exposure and user-side relevance are
\begin{equation}
E_t(w,q)=\sum_{i:\,d_i\in\mathcal D(q,w)}\nu_i,
\qquad
J_t(q)=\sum_i\nu_i\,\rel(q,x_i).
\label{eq:exposure}
\end{equation}
$\mathbf x_t(q)$ is instantiated at two points: the ordered selected websites at website selection and the ordered cited documents at the final answer; \Cref{app:metrics} measures them with W-nDCG@5 and D-nDCG@5. $E_t$ is defined at the final-answer stage, where the exposure reallocation below is derived.
$E_t$ adapts the position-aware impression view of \citet{aggarwal2024geo} to cited documents and normalizes exposure as a share of one answer. The two objective functions over the replayed stream $Q=\bigcup_t Q_t$ are
\begin{equation}
\mathcal O^U=\frac{1}{|Q|}\sum_t\sum_{q\in Q_t}J_t(q),
\qquad
\mathcal O^W(w)=\frac{1}{|Q|}\sum_t\sum_{q\in Q_t}E_t(w,q).
\label{eq:objectives}
\end{equation}
The platform and each website use only their respective realized selection, return, and citation records in later rounds.

\FloatBarrier
\section{Method and baseline details}
\label{app:implementation}

\begin{algorithm}[t]
\caption{\DEC over a request stream}
\label{alg:dec}
\small
\begin{algorithmic}[1]
\Require request stream $\{Q_t\}$, registry $R_1$, provider updater $F_P$, Withhold threshold $m$.
\State Initialize $M^U_0,M^P_0(w)\leftarrow\varnothing$.
\For{each round $t$}
  \State Freeze both histories; initialize $Y_t,Y_t(w)\leftarrow\varnothing$ for all $w$.
  \For{each $q\in Q_t$}
    \State $R_t^C(q)\leftarrow\textsc{Credit}(R_t,C_t(q,\cdot))$; the user agent produces selector order $\sigma_t(q)$ and initial list $L_t^0(q)$ from $R_t^C(q)$.
    \State Compute $\mathcal W_t^-(q)$ and set $L_t(q)\leftarrow\textsc{Withhold}(L_t^0(q),\sigma_t(q),\mathcal W_t^-(q))$; withdrawals hold across all sub-searches of this top-level request, with at most one refill overall.
    \State Serve $L_t(q)$; add the realized market-wide record to $Y_t$ and each selected website's provider-visible record to $Y_t(w)$.
  \EndFor
  \State $M^U_t\leftarrow\operatorname{Append}(M^U_{t-1},Y_t)$; $M^P_t(w)\leftarrow\operatorname{Append}(M^P_{t-1}(w),Y_t(w))$.
  \State If $w$ adapts, $\widehat\rho_{t+1}(w)\leftarrow F_P(\rho_{1:t}(w),M^P_t(w))$; otherwise $\widehat\rho_{t+1}(w)\leftarrow\rho_t(w)$.
  \State All submissions form the next-round registry $R_{t+1}$.
\EndFor
\end{algorithmic}
\end{algorithm}

\subsection{Service chain and history paths}
\label{app:paths}

The service-chain diagnostics record selector choice, document return, citation, and position-weighted final-answer exposure. For the ten adaptive providers we also aggregate selected, returned, and cited counts. Citation records link a provider's service process to its later rewriting.

One top-level request can trigger several sub-searches. Over the $60$-day common trajectory a request runs $1.96$ sub-searches on average, and over a third run at least two. Credit and Withhold both decide from the round-start history of the top-level request and stay consistent within the whole service episode of that request; the withdrawal holds in every sub-search of that request, while at most one refill is granted over the whole episode.

\paragraph{Credit and Withhold.} Credit recalls strictly past main requests by exact match, BM25 and dense retrieval, and takes the union of the three recalls. A recalled pair is kept when the query relation is judged DIRECT or TRANSFERABLE, and the system's own cited titles for that request are joined to it; the action gate requires a support score of $3$ and attaches at most one supporting evidence card per website. Withhold instead recalls strictly past main requests that cited any website in this system's trajectory, taking exact match together with the BM25 and dense top-$12$ lists and keeping the merged top-$20$, and after recall joins the target website's own selection, return and citation records. Its gate requires at least two distinct histories, the dense path to meet the same gate, and the target website never to have been cited; it then keeps the website withdrawn within the current request, with at most one refill using an unselected website that has previously contributed to a final answer. Scored after the run against frozen labels that Withhold never reads, withdrawn seats carry relevance support at about a third of the rate of the adaptive seats it leaves in place ($12.7\%$ versus $36.8\%$), and a refill has the higher website gain about three times as often as the lower, exceeding the withdrawn website by $0.39$ on average (95\% CI $[0.10, 0.67]$). The audit covers every Withhold action on the $60$-day trajectory, across all sub-searches.

Writing $H_t^{\mathrm{wh}}(q)$ for the requests the Withhold
recall returns and $H_{t,\mathrm{dense}}^{\mathrm{wh}}(q)\subseteq H_t^{\mathrm{wh}}(q)$ for those the dense path contributed,
\begin{equation}
\begin{aligned}
\operatorname{Sel}_t(q,w)&=\bigl|\{q'\in H_{t,\mathrm{dense}}^{\mathrm{wh}}(q):w\text{ was selected on }q'\}\bigr|,\\
\operatorname{Cite}_t(q,w)&=\sum_{q'\in H_t^{\mathrm{wh}}(q)}\delta(q',w).
\end{aligned}
\label{eq:withholdcounts}
\end{equation}
with $\delta$ the cited indicator of \Cref{eq:cited} read on the round that served $q'$. The websites
eligible for withholding are
\begin{equation}
\mathcal W_t^-(q)=\{w\in L_t^0(q):\rho_t(w)\ne\rho_1(w),\ \operatorname{Cite}_t(q,w)=0,\ \operatorname{Sel}_t(q,w)\ge m\}.
\label{eq:withholdapp}
\end{equation}
Only complete and consistent joined service states are eligible for the gate; all other cases leave the
current selection unchanged. Both counts use only records in $H_t^{\mathrm{wh}}(q)$, so the test is per type of need,
matching the condition on a topic $c^-$ in \Cref{app:relative-evidence}. Forgetting factors are standard in reputation systems, which discount the past by elapsed time or by transaction frequency \citep{josang2002beta,josang2007survey}. We clear a website's non-delivery counts on a need after $45$ active days without a selection opportunity there; the rule only lifts withholdings. A withdrawal is therefore temporary, while its delivery evidence persists. Both paths are truncated at the
round-start prefix, so same-round and future records are invisible. Website identity takes no part in history recall.

\paragraph{Returning to the registered profile.}\label{app:return-original}Returning to Original changes the active profile but preserves the service evidence used by later actions. Under \Cref{eq:withholdapp}, an exact return removes the revised terms and suspends Withhold. If the website later revises again, comparable non-delivery records enter the gate again. At each eligible update, each content agent uses its own service history to choose among returning to Original, keeping the current profile, and continuing to rewrite. No content agent selected returning to Original. Thus, reversion pauses the revised claims and their selection opportunities while the evidence state remains available to later updates.

\subsection{Provider-side baselines}
\label{app:provider-baselines}

Under Profile Adaptation, the ten adaptive providers continually rewrite their profiles from their own strictly past provider-local service histories. We also compare four ways of writing website profiles adapted from prior work, each submitted once in this experiment: \textbf{GEO Keyword Stuffing} and \textbf{GEO Fluency Optimization} use GEO's keyword and fluency rewrites \citep{aggarwal2024geo}, \textbf{MPMA Authority} uses MPMA's authority phrasing \citep{wang2026mpma}, and \textbf{FedRAG Universal} uses the universal-coverage description from a routing-hijacking attack on ReSLLM \citep{mu2026routinghijack}. All four are placed in website profiles under the same $75$-word limit and selector interface. Rewriting applies to the same ten adaptive providers, and the other ninety profiles are frozen word for word. On the $30$-day common population, we compare their position-weighted exposure at website selection and in the final answer (\Cref{tab:provider-baselines}). Further document-level GEO methods learn generative-engine preferences \citep{wu2025autogeo}, optimize one document for diverse queries \citep{zhou2026ifgeo}, or rewrite for role-augmented user intents \citep{chen2025roleaugmented}; SAGEO and E-GEO evaluate such rewriting in a full search pipeline and in e-commerce \citep{kim2026sageo,bagga2026egeo}, and ranking-incentivized content modification plays the same role in competitive search \citep{goren2020ranking}.

Each comparison profile is submitted once and never revised, covering a provider that registers an optimized profile and leaves it. At registration, this market has no delivery record, so these methods use market-independent patterns. None produces a material exposure gain. Two are roughly level with no rewriting, and the other two are lower at both stages: the share of the ten adaptive providers reaching the selector top five falls from $35.1\%$ to $28.6\%$ and $21.8\%$. Profile Adaptation instead summarizes each website's provider-local service history and raises selection exposure, top-five share, and final-answer exposure. The four comparison methods also keep both quality metrics at or above the no-rewriting level, while Profile Adaptation lowers both.

\textbf{Deviation check over the evaluated provider strategies.} Without platform-side intervention, \Cref{tab:provider-baselines} ranks the same five strategies by final-answer exposure. The two GEO variants are level with no rewriting ($4.23$ and $4.22$), the two injection-style variants fall below it ($3.24$ and $2.76$), and rewriting from provider-local history is the only strategy above it. Among the evaluated provider configurations, $F_P$ gives the provider side the highest payoff, so no deviation within this set is profitable; the other strategies do no better than not rewriting. Profile Adaptation is also the only strategy that costs the user side, by $1.60$ W-nDCG@5 points. The misalignment therefore comes from the best-performing evaluated provider strategy and calls for a selection-time mechanism rather than a fixed text filter (\Cref{app:user-baselines}).

\paragraph{Provider incentives under platform-side action.} The check above holds platform-side policy fixed at no intervention. We now add a stronger deviation and turn that policy on. In an indirect prompt-injection robustness check on a prefix of the request stream \citep{qian2025rankingblindspot,greshake2023indirect,shi2026toolhijacker}, we rank provider strategies by the ten adaptive providers' final-answer exposure, measured against $F_P$ on that same prefix. Without platform-side intervention, explicit injection is both the most profitable strategy available to a provider and the most damaging one: it multiplies their exposure by $2.38$ over $F_P$ while lowering W-nDCG@5 by a further $6.01$ points. Platform-side intervention reverses this ordering (\Cref{app:injection-prompt}): an injected submission is flagged at registration and receives no selections, returns or final-answer uses, so injection now pays less than either alternative in this check, and $F_P$ becomes the best strategy available. Platform-side action therefore removes the profitable deviation and steers the remaining rewriting toward delivery-backed claims (\Cref{sec:better-profiles}), leaving a pair that is best for the provider among the strategies evaluated and better for the user side than no intervention.

\subsection{Platform-side baselines}
\label{app:user-baselines}

Full-history ICL follows the ICL memory policy of Continual Learning Bench \citep{asawa2026clbench}: it adds compact service records from past requests in time order and removes the oldest to meet the shared input limit. Its records carry the same service outcomes that \DEC reads, including whether a returned document entered the final answer and that document's title; \DEC additionally retrieves comparable past requests, Credit verifies them with LLM judgments, and Withhold uses record counts only. It tests direct use of history for website selection; long-context studies show that more input does not always help models use relevant evidence \citep{liu2024lost,du2025context}.

We add three history-based controls under the same profile-selection interface. Global Beta Reputation reranks websites by a Beta$(1,1)$ posterior final-use score over all prior interactions, while Query-conditioned Beta Reputation computes the same score from related prior requests. Relevant-History ICL instead adds matching, strictly prior raw service records to the Selector's context without Credit or Withhold. All three retain Profile Adaptation and the same downstream service chain.

\begin{table}[!h]
\centering
\caption{\textbf{Additional history-based baselines under the $30$-day protocol.}}
\label{tab:query-history}
\small
\setlength{\tabcolsep}{3pt}
\begin{tabular}{@{}lrrrr@{}}
\toprule
System & W-nDCG@5 & D-nDCG@5 & Provider GEO & Welfare Gain \\
\midrule
PA & 51.01 & 41.97 & 5.19 & 0.00 \\
Global Beta Reputation & 41.95 & 39.22 & 2.48 & $-13.31$ \\
Query-conditioned Beta Reputation & 49.17 & 41.01 & \best{7.14} & $+1.20$ \\
Relevant-History ICL & 52.23 & 42.05 & 4.25 & $-0.25$ \\
\textbf{\DEC} & \best{53.27} & \best{43.74} & 5.99 & \best{+3.51} \\
\bottomrule
\end{tabular}
\end{table}
Global Beta Reputation lowers both quality and provider exposure. Conditioning the score on related history gives the highest Provider GEO, but both quality metrics remain below PA. Relevant-History ICL improves W-nDCG@5, while its D-nDCG@5 stays near PA and Provider GEO falls. \DEC achieves the highest W-nDCG@5, D-nDCG@5 and Welfare Gain while raising Provider GEO over PA. Neither a scalar reputation nor raw-history prompting reproduces \DEC's balanced gains for both sides.

\FloatBarrier

VCR is the verifiable-content-reward method of GT-GEO \citep{xu2026mechanism}; it compares successive profiles of the same website and maps its pair-verifiable content reward and suspicion score into the selector order, warning and source policy. Keyword Scrub and Hard Reject are rule-based controls: the former applies a fixed regex filter to profiles, while the latter removes profiles whose suspicion score reaches a threshold before selection. We implement them with GT-GEO's released regex and $0.45$ threshold. The fixed GEO-style phrases targeted by Keyword Scrub occur in seven static profiles and in none of the ten adaptive profiles, whose updates select terms from their own service history. Because the filter weakens only static profiles, the adaptive providers' share of final-answer exposure rises even though none of their own profiles is touched. Each adaptation moves GT-GEO's unit of analysis from document sources to website profiles while keeping its scoring and filtering rules fixed. All methods share the $100$-profile interface, the $50{,}000$-character input cap and one native selector call, and follow their own trajectory from the first change to the selection input or candidate set.

TASR, the defense proposed together with the routing-hijacking attack used in \Cref{app:provider-baselines} \citep{mu2026routinghijack}, multiplies each client's numeric routing score by a trust weight updated from the relevance, profile consistency, and cross-client agreement of returned evidence. Our selector ranks websites listwise without per-website scores, and \DEC conditions on whether returned documents enter the final answer, so TASR is not included as a separate baseline.

Standard per-arm contextual bandits such as LinUCB and Thompson sampling have the same interface mismatch. They convert each website's context into an arm-level value, such as a confidence bound or a posterior reward sample, and choose websites from those values. Applying them here would replace the native listwise selector with a separately defined per-website scoring policy.

A slate or listwise bandit can avoid this decomposition, but it still requires a new policy over website sets and a rule for updating that policy from the selected websites' final-use outcomes. These are alternative selector architectures rather than baselines that operate through the same selector interface. \Cref{app:beta-diagnostic} isolates the exploitation signal of the per-arm family by using a per-website posterior mean to rerank.

\subsection{Joint profile--document optimization}
\label{app:joint-optimization}

We fix documents in the main experiments and focus on website-selection competition before retrieval in the Agentic Web. This is a different stage from the document optimization studied in existing GEO work.

The joint profile--document optimization is also worth studying. We evaluate joint profile--document adaptation under the same provider-local information boundary as the main setting. Rewriting and reindexing all $18.43$ million documents would require corpus-scale LM computation, so we use a retrieval-grounded design: across rounds, each adaptive content agent can revise its profile and multiple documents previously retrieved from its own corpus and returned in past interactions. Both forms of adaptation persist into later rounds, so the provider action space covers the selection-facing profile and the content served after selection.

PA-Joint applies this joint action directly. VCR-Joint extends its version-based mechanism to successive profile and document revisions. \DEC-Joint applies its delivery-evidence mechanism to both forms of adaptation. All three arms use the same provider action space, update timing, and budget. \Cref{tab:joint-optimization} reports the matched comparison.

\section{Supplementary long-horizon results}
\label{app:full60supp}

\subsection{Key paired comparisons}
\label{app:full60}

\begin{table}[!h]
\centering
\caption{\textbf{Paired long-horizon comparisons.} Differences and $95\%$ confidence intervals are in percentage points, except Provider GEO, which is exposure per $100$ requests.}
\label{tab:long-horizon-paired}
\small
\begin{tabular}{llrl}
\toprule
Comparison & Metric & Paired diff. & $95\%$ confidence interval \\
\midrule
Profile Adapt. $-$ Original & W-nDCG@5 & \best{-1.25} & $[-1.83,\ -0.68]$ \\
Credit $-$ Profile Adapt. & W-nDCG@5 & \best{+1.40} & $[0.80,\ 2.06]$ \\
\DEC $-$ Credit & W-nDCG@5 & \best{+0.74} & $[0.20,\ 1.29]$ \\
\DEC $-$ Profile Adapt. & W-nDCG@5 & \best{+2.14} & $[1.44,\ 2.87]$ \\
\DEC $-$ Original & W-nDCG@5 & \best{+0.89} & $[0.18,\ 1.60]$ \\
Profile Adapt. $-$ Original & Provider GEO & \best{+2.49} & $[1.71,\ 3.34]$ \\
\DEC $-$ Original & Provider GEO & \best{+2.23} & $[1.53,\ 3.00]$ \\
\bottomrule
\end{tabular}
\end{table}

All intervals in \Cref{tab:long-horizon-paired} exclude zero. \DEC also improves final-document quality over Profile Adaptation, by $0.70$ D-nDCG@5 points (95\% CI $[0.10, 1.33]$). Request-paired, two-sided tests against Profile Adaptation give $t(4674)=8.87$, $p<.001$ for W-nDCG@5 and $t(4674)=2.58$, $p=.010$ for D-nDCG@5. Across the same paired requests, $10{,}000$ session resamples that re-estimate both anchors and $\lambda$ give $G(\DEC)=+2.01$ points (95\% CI $[+1.32,+2.74]$). \Cref{tab:ablation} in \Cref{sec:ablation} gives the $60$-day levels of the same systems.

\subsection{Equilibrium criteria}
\label{app:equilibrium}

Write $\Theta_t(w)$ for the terms website $w$ has deployed on active day $t$, $\mathcal A$ for the adaptive websites, and $T$ for the final active day. The mean profile residual on day $t$ and the share of changed provider-days in a tail window of length $\tau$ are
\begin{equation*}
\begin{aligned}
\Delta_t^P
&=1-\frac{1}{|\mathcal A|}\sum_{w\in\mathcal A}
 \operatorname{Jaccard}\big(\Theta_t(w),\Theta_{t-1}(w)\big),\\
\chi_{P,\tau}
&=\frac{\big|\{(t,w):T-\tau+1\le t\le T,\ w\in\mathcal A,
 \ \Theta_t(w)\ne\Theta_{t-1}(w)\}\big|}{\tau|\mathcal A|}.
\end{aligned}
\end{equation*}

We empirically identify a trajectory as being in an $(\epsilon_P,\eta_P,\tau)$-approximate rewriting equilibrium over its last $\tau$ active days when
\begin{equation*}
\overline{\Delta}_{P,\tau}
=\frac{1}{\tau}\sum_{t=T-\tau+1}^{T}\Delta_t^P\le\epsilon_P,
\qquad \chi_{P,\tau}\le\eta_P.
\end{equation*}
At $\tau=20$ both arms satisfy $(\epsilon_P,\eta_P)=(0.02,0.05)$: over active days $41$ to $60$ Profile Adaptation has $\overline{\Delta}_{P,20}=0.0151$ and changes a deployed term set on $2.5\%$ of provider-days, and \DEC $0.0143$ and $2.0\%$. Rewriting is far from stationary earlier on: over the first third of the trajectory, Profile Adaptation has mean residual $0.22$ and changes a deployed term set on $38.9\%$ of provider-days, and \DEC $0.21$ and $39.2\%$. This invariance under repeated updates is the stability notion of \citet{perdomo2020performative}, whose stable points need not be optimal. The same slot limit and candidate construction allow the high early update rates just reported, and the two arms settle on different claim compositions (\Cref{fig:unbacked-claims}), so the stable point depends on the platform-side policy rather than on the slot limit.

Over the same tail window, the complete profile text is identical to the previous day on $93.0\%$ of Profile Adaptation provider-days and $94.5\%$ of \DEC provider-days; the remaining text changes with an unchanged term set only reorder existing terms.

\paragraph{Scope of the criterion.} This empirical equilibrium criterion measures profile stability under repeated $F_P$ updates. The second continual updater in \Cref{app:updater-robustness} tests robustness for the full provider side. Changing one website's updater while holding its competitors' strategies fixed is instead an individual-provider Nash evaluation. The platform-side condition below likewise compares only the evaluated class $\Pi_U^{\mathrm{eval}}$.

The platform-side condition uses the same tail window, applied at the ranking point the platform-operated user agent controls directly. Let $\Pi_U^{\mathrm{eval}}=\{\pi_U^{\mathrm{base}},\pi_U^{\mathrm{Credit}},\pi_U^{\DEC}\}$ be the three evaluated policies run with the same provider updater, where $\pi_U^{\mathrm{Credit}}$ uses Credit without Withhold. Comparing policies within an evaluated class follows empirical game-theoretic analysis, which reasons about simulated games over restricted strategy sets \citep{wellman2006methods}, and empirical mechanism design, which chooses mechanisms by their simulated outcomes \citep{vorobeychik2006empirical,viqueira2019empirical}. Let $\widehat S^W_{U,\tau}(\pi_U;F_P)$ be tail-window W-nDCG@5 along the complete trajectory produced by $\pi_U$. For $\pi_U\in\Pi_U^{\mathrm{eval}}$, its within-class selection-policy gap is
\begin{equation*}
\widehat{\operatorname{Gap}}^W_{U,\tau}(\pi_U;\Pi_U^{\mathrm{eval}})
=\max_{\pi'_U\in\Pi_U^{\mathrm{eval}}}
\left(\widehat S^W_{U,\tau}(\pi'_U;F_P)-\widehat S^W_{U,\tau}(\pi_U;F_P)\right).
\end{equation*}

The two criteria answer separate questions. The rewriting criterion asks whether profiles have stabilized under $F_P$, while the selection-policy gap ranks the three evaluated user-agent policies by tail-window website-selection quality. The provider-side statistics above place \DEC in an approximate rewriting equilibrium. Over active days $41$--$60$, Profile Adaptation, Credit and \DEC obtain W-nDCG@5 values of $49.46$, $50.79$ and $52.27$, giving gaps of $2.81$, $1.48$ and $0.00$ points. Thus, \DEC meets the provider-side stability criterion, while its user-agent selection policy ranks first within $\Pi_U^{\mathrm{eval}}$.

\FloatBarrier
\section{Mechanism evidence and ecosystem responses}

\subsection{Service conversion and profile evidence}
\label{app:funnel}

Compared with Profile Adapt., \DEC has $5.6\%$ fewer selected episodes and $3.4\%$ less final-answer exposure for the ten adaptive providers, while retaining $89.7\%$ of Profile Adaptation's exposure gain over Original. Cited episodes and cited documents remain nearly unchanged, so exposure per selection increases (\Cref{tab:service-funnel}). Citation rate rises by $2.73$ percentage points, and cited documents per selected episode by $6.4\%$. An offline audit further shows that the share of adaptive providers' final-answer exposure on irrelevant matches falls by $14.6\%$ relative to Profile Adapt. Static-Web exposure is reported separately in \Cref{app:market-structure}.

\begin{table}[h]
\centering
\caption{\textbf{Adaptive-provider service funnel and conversion.} Counts, rates, and exposure cover only the ten adaptive providers in the shared long-horizon population.}
\label{tab:service-funnel}
\scriptsize
\setlength{\tabcolsep}{4pt}
\begin{tabular}{lrrrr}
\toprule
Service quantity & Original & Profile Adapt. & Credit & \DEC \\
\midrule
Selected episodes & 2{,}095 & 2{,}848 & 2{,}692 & 2{,}688 \\
Cited episodes & 1{,}000 & 1{,}412 & 1{,}305 & 1{,}406 \\
Cited documents & 1{,}993 & 2{,}763 & 2{,}573 & 2{,}775 \\
Final-answer exposure & 239.16 & 355.44 & 297.52 & 343.50 \\
Exposure per selection & 0.1142 & 0.1248 & 0.1105 & 0.1278 \\
Citation rate & $47.73\%$ & $49.58\%$ & $48.48\%$ & $52.31\%$ \\
Cited documents per selected episode & 0.95 & 0.97 & 0.96 & 1.03 \\
\bottomrule
\end{tabular}
\end{table}

\paragraph{Delivery-backed profile terms.}
\label{app:profiles}
Historical delivery backing is measured over rounds preceding the current profile update. A deployed term is backed when an earlier request expresses the need described by that term and at least one document returned by the same website for that request is retained in the final answer. The first condition uses normalized matching between profile terms and request text; the second uses returned-document and final-citation IDs. \Cref{fig:unbacked-claims} reports the resulting endpoint composition and its trajectory.

\subsubsection{Profile replacement under a fixed selector}
\label{app:profile-replacement}

To test whether delivery-backed profiles by themselves improve website selection, we fix the selector and the requests and replace only the profiles of the ten adaptive websites. Each request uses the profiles that Profile Adapt. or \DEC deployed at the start of its round; website selection reads only these profiles, with no Credit evidence and no Withhold. Consistent with \Cref{fig:unbacked-claims}, the comparison covers active days $21$--$60$, the last $40$ of the $60$ active days. Weighted by request, \DEC lowers the share of unbacked claims from $21.4\%$ to $14.5\%$. We run the selector three times for each profile set and average the runs. With \DEC's profiles, W-nDCG@5 rises by $2.47$ points ($95\%$ confidence interval $[1.47, 3.50]$), and the share of relevant matches among adaptive websites' Top-5 selections rises from $31.6\%$ to $39.3\%$. The gain holds on the $404$ requests on which \DEC's own history matching accepts no earlier request, where W-nDCG@5 rises by $2.27$ points ($95\%$ confidence interval $[0.88, 3.72]$). The effect therefore does not require a reusable record of the current need: it also reaches needs without one through the profiles the mechanism has already shaped.

\subsubsection{A fixed-candidate reputation diagnostic}
\label{app:beta-diagnostic}

This diagnostic tests whether \DEC's gain can be reproduced by reducing each website's related delivery history to one reputation score. Selecting sources from recorded outcomes has precedents in federated search, which uses query click-through data for resource selection \citep{arguello2009classification}, and in multi-platform agent markets, where the user agent keeps records of whether each platform's earlier claims were borne out \citep{hong2026platformscompete}. For a controlled comparison, we keep each completed trajectory and its candidate set fixed and rerank the same candidates using only that score. For every request, the score is the Beta$(1,1)$ posterior mean of a website's final-use rate on related earlier requests; retrieval of those requests does not use website identity. We then apply the same selection cut and provider-family cap.

Applied to the Profile Adaptation trajectory, this reranker lowers W-nDCG@5 by $2.24$ points (95\% CI $[-3.74,-0.79]$); applied to \DEC's trajectory, it lowers W-nDCG@5 by $3.68$ points ($[-5.17,-2.25]$). Both intervals pair outcomes by request and resample semantic sessions. Because each trajectory has its own candidate sets and platform states, the two drops should be read separately rather than as a head-to-head difference. In this setting, reducing related outcomes to one score loses information needed for website matching. Recent work likewise models trust by skill when agents have heterogeneous capabilities and evidence is sparse \citep{xia2026conditional}. \DEC instead keeps delivery evidence tied to the need it supports: Credit presents verified past delivery to the selector, while Withhold acts on repeated non-delivery.

\subsubsection{Request-level mechanism cases}
\label{app:mechanism-cases}

Three request-level traces make the aggregate evidence above concrete.

\paragraph{Delivery-grounded Rewrite.} An Academia.edu profile update replaces \texttt{homepage} and \texttt{profile} with \texttt{journalist} and \texttt{sustainability}. The old terms had no delivery backing in the preceding two intervals, while both replacements had prior final-use evidence. In the next interval, requests about journalist credibility, sustainable manufacturing, and product lifecycle management again use Academia.edu documents.

Across the complete trajectories, \DEC records 65 more Academia.edu selections with frozen relevance support, 22 fewer without it, and an 11.9-point higher citation rate than Profile Adapt. (\Cref{app:topic-heterogeneity}). At the shared endpoint, Profile Adapt. still advertises \texttt{pdf} and \texttt{doc}: earlier \texttt{pdf} matches include cases where the content agent withheld its own return, while \texttt{doc} matches no selected request, and neither has a matching document used in a final answer. All eight \DEC terms have historical final-use support, including \texttt{sustainability} and \texttt{plm} through the site's \textit{The Role of Additive Manufacturing in Improving Resource Efficiency and Sustainability} and \textit{What is Product Lifecycle Management (PLM)}. The update therefore moves limited profile slots from generic or unsupported terms to needs for which the website has demonstrated delivery ability.

\paragraph{Credit promotes a website with prior delivery.} For the request \textit{why are cubans protesting}, Credit attaches a supporting evidence card recording that the New York Times article \textit{How Hunger Has Fueled the Cuba Protests} entered an earlier final answer for the same need. The Top-5 website set remains unchanged, while the New York Times moves from rank 4 to rank 1; the current output then uses three documents from the website, including two grade-3 documents, and W-nDCG@5 rises by $23.79$ points. This trace connects verified past delivery to renewed high-relevance delivery by the same website.

\paragraph{Withhold reallocates an unconverted opportunity.} For the request \textit{fjord formation process}, WordHippo appears relevant from its profile terms, but none of the documents it returned in four comparable past contacts entered the final answer. Withhold removes WordHippo at rank 4 and backfills ScienceDirect from rank 6 in the same Selector ranking. ScienceDirect then supplies the grade-2 document \textit{Fjords as Aquatic Critical Zones}, which enters the current final answer and raises W-nDCG@5 by $5.79$ points. This trace uses recorded delivery outcomes to give a scarce opportunity to a website that provides an answer-bearing document for the current request.

The cases show that delivery outcomes shape both the capabilities content agents claim in later profiles and the user agent's allocation of scarce website-selection opportunities in the current request.

\subsection{Website matching and market structure}
\label{app:market-structure}

All market-level analyses use the shared long-horizon population. A selection is one website served in the Top-5 of a request's first Selector search, after any Withhold reallocation. A request--website match has frozen relevance support when its website-level gain in W-nDCG@5 is positive.

\begin{figure*}[t]
\centering
\includegraphics[width=\textwidth]{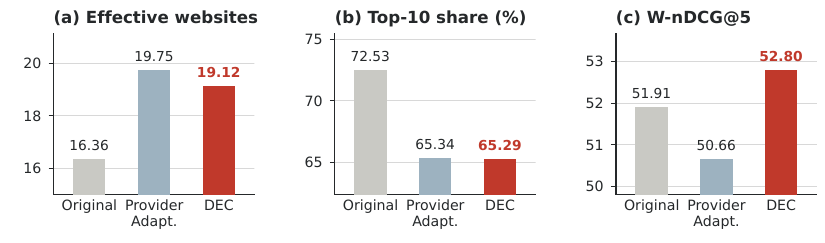}
\caption{\textbf{Market openness and website-selection quality.} (a) Effective number of websites; (b) selection share of the ten most frequent websites; (c) W-nDCG@5.}
\label{fig:appendix-market-structure}
\end{figure*}

We measure market openness from this served-selection distribution (\Cref{fig:appendix-market-structure}). The effective number of websites is its inverse Herfindahl index, and Top-10 share is the fraction of selection opportunities assigned to the ten most frequent websites. Profile Adapt. broadens participation; \DEC preserves that broader structure while improving website-selection quality.

The numerator of W-nDCG@5 separates additively into contributions from adaptive websites and the static web. Under Profile Adapt., the adaptive gain is smaller than the static loss; \DEC raises both contributions.

\begin{table}[t]
\centering
\caption{\textbf{Source contributions to W-nDCG@5.} Values are system contrasts in percentage points.}
\label{tab:source-contributions}
\small
\begin{tabular}{lrrr}
\toprule
System contrast & Adaptive & Static web & Total W@5 \\
\midrule
Profile Adapt. $-$ Original & $+0.72$ & $-1.97$ & $-1.25$ \\
\DEC $-$ Profile Adapt. & $+0.29$ & $+1.85$ & $+2.14$ \\
\bottomrule
\end{tabular}
\end{table}

Splitting requests by whether at least one adaptive website has frozen relevance support locates both the loss and the recovery. When such a website exists, the adaptive gain under Profile Adapt. nearly offsets the static loss. Otherwise, the static web accounts for the entire loss in that group.

\begin{table*}[t]
\centering
\caption{\textbf{W-nDCG@5 source contributions by adaptive supply.} PA denotes Profile Adapt.; values are percentage-point contrasts.}
\label{tab:supply-contributions}
\scriptsize
\setlength{\tabcolsep}{3pt}
\begin{tabular}{lrrrrrrr}
\toprule
Adaptive supply & Requests & \multicolumn{3}{c}{PA $-$ Original} & \multicolumn{3}{c}{\DEC $-$ PA} \\
\cmidrule(lr){3-5}\cmidrule(lr){6-8}
condition & (share) & Adapt. & Static & Total & Adapt. & Static & Total \\
\midrule
No supported adaptive site & 2{,}250 ($48.1\%$) & 0.00 & $-2.66$ & $-2.66$ & 0.00 & $+1.89$ & $+1.89$ \\
$\ge1$ supported adaptive site & 2{,}425 ($51.9\%$) & $+1.38$ & $-1.33$ & $+0.05$ & $+0.56$ & $+1.82$ & $+2.37$ \\
\bottomrule
\end{tabular}
\end{table*}

On requests with no relevance-supported adaptive website, \DEC narrows the W-nDCG@5 gap from Original from $2.66$ to $0.77$ points, repairing $71\%$ of the quality loss.

Counted as first-slate top-5 seats, Profile Adapt. moves $801$ seats away from the static web relative to Original, and \DEC returns $151$ of them, recovering $94.2\%$ of the static-web quality-contribution loss while further increasing the adaptive contribution. Static-Web final-answer exposure and cited-document volume remain nearly unchanged between Profile Adaptation and \DEC. \DEC's gain therefore comes from better website selection, mostly through the static-Web contribution (\Cref{tab:source-contributions}), rather than from moving exposure away from the static web.

Grouping websites by their net change in total selections over the complete trajectory gives the same direction: the opportunities gained by net-inflow websites have a $28.1$-point higher support rate than the opportunities lost by net-outflow websites. This website-level net-flow analysis summarizes changes across website groups rather than tracing one-for-one replacements within requests.

\subsection{Topic and website heterogeneity}
\label{app:topic-heterogeneity}

Topic membership is assigned non-exclusively from the sectors of the highest-grade sources in the frozen qrels, so one request can enter multiple topics. The table reports three representative markets and three additional markets. Selection changes are \DEC minus Profile Adaptation per $100$ requests in that topic; W-nDCG@5 and support-share changes are percentage points. Here, ``supp.'' and ``unsupp.'' denote selections with and without frozen relevance support. Under the evaluated adaptive websites and request stream, General Reference, News, and Technology show limited, mixed, and relatively abundant relevance-supported adaptive supply.

\begin{table*}[t]
\centering
\caption{\textbf{Topic-level quality and selection changes.} W@5 and support-share changes are percentage points; selection changes are per $100$ topic requests.}
\label{tab:topic-changes}
\small
\setlength{\tabcolsep}{3pt}
\begin{tabular}{lrrrrrr}
\toprule
Topic & PA$-$O W@5 & \DEC$-$PA W@5 & $\Delta$ total & $\Delta$ supp. & $\Delta$ unsupp. & $\Delta$ supp. share \\
\midrule
General reference & $-2.48$ & $+3.15$ & $-7.52$ & $+1.02$ & $-8.54$ & $+8.5$pp \\
Language reference & $-4.71$ & $+7.31$ & $-20.18$ & $-1.32$ & $-18.86$ & $+9.2$pp \\
News & $-0.25$ & $+2.65$ & $-1.43$ & $+3.03$ & $-4.46$ & $+11.4$pp \\
Research & $-0.76$ & $+2.59$ & $+1.28$ & $+1.39$ & $-0.11$ & $+2.1$pp \\
Technology & $+0.62$ & $+3.95$ & $+11.19$ & $+15.07$ & $-3.88$ & $+9.9$pp \\
Essays \& blogs & $+1.92$ & $+3.35$ & $+2.32$ & $+3.76$ & $-1.45$ & $+2.9$pp \\
\bottomrule
\end{tabular}
\end{table*}

The three representative markets in \Cref{tab:topic-changes} show how \DEC responds to different levels of relevance-supported adaptive supply. General Reference already has high-contribution static sources but relatively few supported adaptive matches, so \DEC reduces unsupported adaptive matches and restores displaced reference sources. News contains both supported and unsupported adaptive matches, so \DEC mainly changes participation composition: total adaptive participation changes little, while its supported share rises by $11.4$ points. Technology has more relevance-supported adaptive websites, and \DEC expands their participation.

For each provider and topic, the provider--topic breadth analysis uses the same symmetric request set: the union of requests on which that provider enters the served Top-5 under Profile Adaptation or \DEC. Both outcomes are paired on this union. \Cref{fig:robustness} shows the larger topics and the corresponding provider marginals. Topic aggregates describe complete topic markets, while the provider--topic request sets test whether improvement extends across active markets. \DEC raises W-nDCG@5 in nearly all of these markets and repairs nearly all markets harmed by Profile Adaptation. This breadth remains stable across the tested minimum cell sizes.

Using the shared long-horizon population and Profile Adaptation as the baseline, the website cases show two responses to recorded delivery histories: supported expansion and clearer specialization. When a website's recorded deliveries span more kinds of requests, its later opportunities can broaden; when they concentrate on a narrower range of needs, its opportunities can concentrate there. Academia follows the first: its supported selections rise while its unsupported ones fall, so the traffic it gains lands on needs it has already served. wikiHow follows the second: it loses selections overall, yet its supported selections still rise, because the loss falls almost entirely on unsupported ones. Both end with a higher supported share and a higher citation rate.

\section{Robustness and sensitivity}
\label{app:uncertainty}
\label{app:additional-robustness}

We evaluate statistical uncertainty and execution robustness separately. For systems sharing a request population, we compute paired differences within requests and resample the semantic sessions defined in \Cref{app:data-views}. The $60$- and $30$-day populations contain $1{,}332$ and $626$ sessions, respectively. All intervals over WildRegistry trajectories use $10{,}000$ session-bootstrap draws with seed $20260917$ and condition on the realized trajectories \citep{efron1993bootstrap,koehn2004statistical}. This avoids treating requests from the same session as independent. The external label-validation audit in \Cref{app:judgements} evaluates a separate population of benchmark episodes and therefore resamples episodes. Completed runs varying selector execution, model backbone, adaptive-website composition and horizon give separate checks of execution robustness. The independent full replays below regenerate selections, provider-local histories and profile updates to test whether the long-run result recurs across different realized adaptation trajectories.

\paragraph{Stability under execution perturbations.} Fixed-selector profile replacement is positive in all three repeats, with a between-run standard deviation of only $0.29$ points (\Cref{app:funnel}). \DEC's W-nDCG@5 gain over Profile Adaptation is also positive, with confidence intervals excluding zero, across all three backbones, the second adaptive-website cohort and the $60$-day protocol (\Cref{app:full60,app:thirtyday,app:alternative-adaptive-websites}). The website-selection gain is therefore not tied to one selector call, model family, website cohort or horizon.

\subsection{Stability across independent executions}
\label{app:across-run}

Independent full replays reproduce \DEC's long-run result despite different early histories. We repeat the complete $60$-day Profile Adaptation--\DEC interaction in two execution periods with the same request stream and configuration. During the first ten active days, $36.6\%$ of requests select a different set of adaptive websites across the two \DEC runs, and $19.5\%$ use a different set in the final answer. These differences establish two distinct realized adaptation paths.

\begin{table}[t]
\centering
\caption{\textbf{Allocation variation across independent \DEC runs.} Values are total-variation gaps across the ten adaptive websites.}
\label{tab:run-allocation-gap}
\small
\setlength{\tabcolsep}{5pt}
\begin{tabular}{lrr}
\toprule
Allocation & First 10 days & Final 10 days \\
\midrule
Website-selection opportunities & $25.7\%$ & $4.3\%$ \\
Final use & $20.0\%$ & $3.9\%$ \\
Provider GEO & $21.7\%$ & $3.0\%$ \\
\bottomrule
\end{tabular}
\end{table}

All three allocation gaps contract by the final ten days. The Provider GEO gap falls from $21.7\%$ to $3.0\%$, about one seventh of its initial size, and website-selection and final-use gaps follow the same pattern. The two executions therefore follow different early paths, yet their \DEC allocations become closely aligned late in the run.

A same-website comparison gives the same pattern. Answers.com records $30$ successful deliveries in Replay 2 and $13$ in Replay 1 during the first ten days, but its final-ten-day exposure shares are close, at $3.3\%$ in Replay 2 and $4.5\%$ in Replay 1. The larger early delivery count in Replay 2 therefore does not persist as a larger late-run exposure share for this website.

\begin{table}[t]
\centering
\caption{\textbf{Complete outcomes in two independent $60$-day replays.}}
\label{tab:independent-runs}
\small
\setlength{\tabcolsep}{5pt}
\begin{tabular}{llrrr}
\toprule
Replay & System & W-nDCG@5 & D-nDCG@5 & Provider GEO \\
\midrule
\multirow{2}{*}{1} & Profile Adapt. & 51.51 & 40.75 & 7.39 \\
& \DEC & 52.89 & 42.06 & 7.01 \\
\midrule
\multirow{2}{*}{2} & Profile Adapt. & 50.76 & 40.56 & 7.89 \\
& \DEC & 52.76 & 41.82 & 6.87 \\
\bottomrule
\end{tabular}
\end{table}
\DEC improves both quality measures over Profile Adaptation in each execution. Together, the two replays show that \DEC's quality advantage recurs even when execution variation changes the early history, while the resulting allocation paths become closely aligned by the end of the run.

\subsection{Robustness to a less restricted profile updater}
\label{app:updater-robustness}

\DEC's long-run quality and profile-composition advantages persist with a less restricted profile updater. We rerun the complete $60$-day trajectories after removing the explicit limit of eight candidate terms while retaining the common $75$-word profile budget. Profile Adaptation and \DEC deploy more than eight terms on $18.1\%$ and $4.7\%$ of provider-days, confirming that the added rewriting space changes realized profiles.

\begin{table}[!h]
\centering
\caption{\textbf{Complete $60$-day results with the second profile updater.}}
\label{tab:second-updater}
\small
\setlength{\tabcolsep}{4pt}
\begin{tabular}{lrrrr}
\toprule
System & W-nDCG@5 & D-nDCG@5 & Provider GEO & Unbacked claims \\
\midrule
\emph{Original} & 51.91 & 41.32 & 5.11 & -- \\
Profile Adapt. & 49.68 & 38.98 & 6.89 & $30.8\%$ \\
\DEC & 51.21 & 40.45 & 7.09 & $15.7\%$ \\
\DEC $-$ PA & $+1.53\ [1.02,\ 2.03]$ & $+1.46\ [0.89,\ 2.04]$ & $+0.20\ [-0.24,\ 0.63]$ & $-49\%$ \\
\bottomrule
\end{tabular}
\end{table}

Relative to Profile Adaptation, \DEC reduces unbacked claims from $30.8\%$ to $15.7\%$ and improves both user-side quality measures with confidence intervals excluding zero. Both systems remain above Original in Provider GEO, while their $+0.20$ difference is not distinguishable from zero. Over the final $20$ active days, $91.5\%$ and $91.0\%$ of their deployed profiles retain the same term set. Removing the term limit therefore changes realized profiles but preserves stable rewriting and \DEC's quality and profile-composition advantages.

\subsection{Heterogeneity by request difficulty}
\label{app:distribution}

We group requests by the number of documents with grade at least $2$ in the frozen candidate pool: at most $2$, $3$--$7$, and more than $7$.

\begin{table}[t]
\centering
\caption{\textbf{W-nDCG@5 differences by request difficulty.} Values are paired differences with $95\%$ confidence intervals; parentheses give each group's share of requests.}
\label{tab:request-difficulty}
\small
\begin{tabular}{llrl}
\toprule
High-grade documents & Comparison & Paired diff. & $95\%$ confidence interval \\
\midrule
$\le 2$ ($53.9\%$) & \DEC $-$ PA & \best{+2.23} & $[1.28,\ 3.26]$ \\
 & Credit $-$ PA & $+1.23$ & $[0.29,\ 2.27]$ \\
 & \DEC $-$ Credit & $+1.00$ & $[0.27,\ 1.75]$ \\
\midrule
$3$--$7$ ($22.9\%$) & \DEC $-$ PA & \best{+2.32} & $[1.00,\ 3.80]$ \\
 & Credit $-$ PA & $+1.87$ & $[0.91,\ 2.85]$ \\
 & \DEC $-$ Credit & $+0.45$ & $[-0.71,\ 1.74]$ \\
\midrule
$>7$ ($23.2\%$) & \DEC $-$ PA & \best{+1.77} & $[0.59,\ 3.02]$ \\
 & Credit $-$ PA & $+1.34$ & $[0.28,\ 2.41]$ \\
 & \DEC $-$ Credit & $+0.43$ & $[-0.37,\ 1.26]$ \\
\bottomrule
\end{tabular}
\end{table}

The metric in \Cref{tab:request-difficulty} is W-nDCG@5 in percentage points. All three \DEC $-$ Profile Adaptation confidence intervals exclude zero. Shuffling the three group labels $2{,}000$ times gives $p=0.68$ for the between-group range of \DEC $-$ Profile Adaptation and $p=0.56$ for \DEC $-$ Credit, revealing no detectable difference among the bins.

\subsection{Weight sensitivity}
\label{app:weights}
\begin{figure}[t]
\centering
\includegraphics[width=\linewidth]{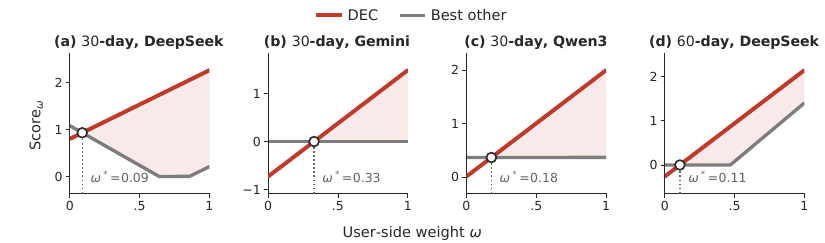}
\caption{\textbf{Weight sensitivity without the Original anchor.} Each system is scored as $\omega\Delta S+(1-\omega)\Delta V$, its W-nDCG@5 and Provider GEO differences from Profile Adaptation, so the reference system is the flat line at zero and $\omega$ is how much of the weight goes to the user side. Grey is the highest-scoring system other than \DEC at that weight; the shaded range is where \DEC leads, starting at $\omega^*$. Equal weighting, $\omega=0.5$, lies inside all four ranges.}
\label{fig:weight-ranges}
\end{figure}

The calibrated $\lambda$ fixes one exchange rate; this subsection checks the ordering without it. Let $\omega\in(0,1]$ be the user-side weight in
\begin{equation}
\operatorname{Score}_\omega(s)=\omega\,\Delta S_s+(1-\omega)\,\Delta V_s,
\end{equation}
where $\Delta S_s$ and $\Delta V_s$ are system $s$'s W-nDCG@5 and Provider GEO differences from Profile Adaptation, so the reference system scores $0$ at every weight. This form needs no second anchor and applies to every backbone. The crossover weight $\omega^*$ is $0.089$ for DeepSeek at $30$ days, $0.178$ for Qwen, $0.329$ for Gemini and $0.108$ for DeepSeek at $60$ days (\Cref{fig:weight-ranges}). The largest value is $0.329$, so one overall condition is enough: when user-side quality receives at least $32.9\%$ of the weight, \DEC ranks first in every setting. Equal weighting assigns $50\%$ to user-side quality and meets this condition \citep{xu2026mechanism}. The threshold is stated in W-nDCG@5 points and Provider GEO per $100$ requests. Gemini sets the strictest threshold because \DEC retains rather than increases exposure there ($\Delta V=-0.73$), so giving more weight to exposure works against it.

\subsection{Paired uncertainty of the 30-day comparisons}
\label{app:thirtyday}

The table reports the key comparisons from \Cref{sec:main,sec:robustness}. Differences are paired within each request, and the confidence intervals resample the $626$ semantic sessions.

\begin{table}[t]
\centering
\caption{Paired $95\%$ confidence intervals for the $30$-day comparisons.}
\label{tab:thirtyday-ci}
\small
\begin{tabular}{llrl}
\toprule
Comparison & Metric & Paired diff. & $95\%$ confidence interval \\
\midrule
\DEC $-$ Full-history ICL & W-nDCG@5 & \best{+7.16} & $[5.18,\ 9.19]$ \\
\DEC $-$ Full-history ICL & D-nDCG@5 & \best{+3.35} & $[2.05,\ 4.67]$ \\
Query-cond. History $-$ PA & W-nDCG@5 & \best{+1.22} & $[0.32,\ 2.10]$ \\
Query-cond. History $-$ PA & D-nDCG@5 & $+0.08$ & $[-0.93,\ 1.10]$ \\
Query-cond. History $-$ PA & Provider GEO & $-0.94$ & $[-1.60,\ -0.31]$ \\
\DEC $-$ Query-cond. History & W-nDCG@5 & \best{+1.04} & $[0.21,\ 1.88]$ \\
\DEC $-$ Query-cond. History & D-nDCG@5 & \best{+1.69} & $[0.67,\ 2.70]$ \\
\DEC $-$ Query-cond. History & Provider GEO & \best{+1.74} & $[1.00,\ 2.50]$ \\
\DEC $-$ Profile Adapt. (DeepSeek) & W-nDCG@5 & \best{+2.26} & $[1.28,\ 3.33]$ \\
\DEC $-$ Profile Adapt. (Gemini) & W-nDCG@5 & \best{+1.49} & $[0.48,\ 2.52]$ \\
\DEC $-$ Profile Adapt. (Qwen) & W-nDCG@5 & \best{+2.00} & $[0.94,\ 3.07]$ \\
\DEC $-$ Profile Adapt. (DeepSeek) & D-nDCG@5 & \best{+1.76} & $[0.90,\ 2.65]$ \\
\DEC $-$ Profile Adapt. (Gemini) & D-nDCG@5 & $+0.02$ & $[-0.85,\ 0.91]$ \\
\DEC $-$ Profile Adapt. (Qwen) & D-nDCG@5 & \best{+1.19} & $[0.46,\ 1.94]$ \\
\bottomrule
\end{tabular}
\end{table}

\begin{table}[t]
\centering
\caption{\textbf{Request-paired, two-sided $t$-tests for the DeepSeek $30$-day comparisons.} Each cell reports the paired difference $\Delta=$ \DEC $-$ baseline in percentage points, followed by $t$ and $p$. All tests use $1{,}465$ requests ($df=1464$).}
\label{tab:thirtyday-ttests}
\footnotesize
\setlength{\tabcolsep}{2.4pt}
\begin{tabular}{@{}lcc@{}}
\toprule
Baseline & W-nDCG@5 $(\Delta;\ t;\ p)$ & D-nDCG@5 $(\Delta;\ t;\ p)$ \\
\midrule
Profile Adaptation & $+2.26;\ 5.31;\ <.001$ & $+1.76;\ 3.74;\ <.001$ \\
Full-history ICL & $+7.16;\ 11.15;\ <.001$ & $+3.35;\ 5.77;\ <.001$ \\
VCR & $+2.06;\ 4.48;\ <.001$ & $+1.76;\ 3.58;\ <.001$ \\
Keyword Scrub & $+2.87;\ 6.69;\ <.001$ & $+1.82;\ 3.72;\ <.001$ \\
Hard Reject & $+2.84;\ 6.60;\ <.001$ & $+2.10;\ 4.24;\ <.001$ \\
Global Beta Reputation & $+11.32;\ 18.01;\ <.001$ & $+4.52;\ 7.67;\ <.001$ \\
Query-cond. Beta Reputation & $+4.10;\ 7.21;\ <.001$ & $+2.73;\ 4.38;\ <.001$ \\
Relevant-History ICL & $+1.04;\ 2.42;\ .016$ & $+1.69;\ 3.23;\ .001$ \\
\bottomrule
\end{tabular}
\end{table}

In \Cref{tab:thirtyday-ci}, differences are in percentage points; the website-selection intervals exclude zero under all three backbones. Request-paired t-tests in \Cref{tab:thirtyday-ttests} show that \DEC outperforms all eight baselines in both W-nDCG@5 and D-nDCG@5.
\subsection{Alternative adaptive websites}
\label{app:alternative-adaptive-websites}

To test sensitivity to target identity, we repeat the $30$-day comparison on the same $1{,}465$ requests, with the ten adaptive websites replaced by ranks $11$--$20$ of the same preregistered ordering. That ordering is fixed by pre-simulation selection, return and citation histories. The request stream, $100$-site registry, provider updater and platform-side policies are unchanged. \Cref{tab:alt-adaptive} gives the endpoint values. Relative to Profile Adaptation, \DEC improves W-nDCG@5 by $2.03$ points ($95\%$ CI $[1.00,\ 3.14]$) and D-nDCG@5 by $2.04$ points ($[0.95,\ 3.21]$), and retains $93.3\%$ of its position-weighted exposure gain over Original.

We next vary the adaptive share by increasing the population from ten websites ($10\%$) to thirty ($30\%$). In this broader setting, \DEC improves W-nDCG@5 by $3.34$ points and D-nDCG@5 by $0.47$ points over Profile Adaptation. Its Provider GEO remains $2.08$ above Original, yielding a Welfare Gain of $+1.91$ (\Cref{tab:adaptive-populations}).

\begin{table}[h]
\centering
\caption{\textbf{Robustness to adaptive-website identity and population size under the $30$-day protocol.} Bold marks the better value within each population.}
\label{tab:alt-adaptive}
\label{tab:adaptive-populations}
\small
\setlength{\tabcolsep}{3.5pt}
\begin{tabular}{@{}llrrrr@{}}
\toprule
Population & System & W-nDCG@5 & D-nDCG@5 & Provider GEO & Welfare Gain \\
\midrule
\multirow{2}{*}{Alternative 10} & \emph{Profile Adapt.} & 50.63 & 40.29 & \best{2.64} & 0.00 \\
& \textbf{\DEC} & \best{52.66} & \best{42.32} & 2.59 & \best{+1.90} \\
\midrule
\multirow{2}{*}{Random 30} & \emph{Profile Adapt.} & 50.15 & 41.26 & \best{12.29} & 0.00 \\
& \textbf{\DEC} & \best{53.49} & \best{41.73} & 9.38 & \best{+1.91} \\
\bottomrule
\end{tabular}
\end{table}
Across both changes, \DEC improves W-nDCG@5 and D-nDCG@5 over Profile Adaptation and attains positive Welfare Gain. The two-sided result therefore holds for another adaptive group and when the adaptive share rises from $10\%$ to $30\%$.

\FloatBarrier
\section{Additional formal analysis}
\label{app:proofs}

This section derives the relative delivery-evidence result used in the mechanism analysis in \Cref{sec:analysis} and the long-horizon analysis in \Cref{sec:why}.

\subsection{Relative delivery evidence}
\label{app:relative-evidence}

Each actual selection generates a provider-visible record that enters the website's provider-local service history. While the request-specific Withhold condition continues to hold, topics with repeated non-delivery stop gaining records, whereas topics whose documents are cited continue to earn them. To formalize this provider-side pressure, let $h_t^+(w,c)$ and $h_t^0(w,c)$ count website $w$'s past selections on topic $c$ with and without a citation, respectively. Define
\begin{equation}
\ell_t(w,c)=h_t^+(w,c)-\gamma\,h_t^0(w,c),
\qquad \gamma>0.
\label{eq:ledger}
\end{equation}
Here $\gamma$ weights non-delivery. The deployed updater reads the complete profile trajectory and provider-local service history in \Cref{app:prompts}; $\ell_t$ summarizes the delivery evidence in those inputs: the larger $\ell_t(w,c)$ is, the stronger the historical support for keeping topic $c$ in a profile of limited capacity.

\begin{lemma}[Relative delivery-evidence advantage over a continued-withholding interval]
\label{prop:alignment}
Suppose that by round $t$, website $w$ has been selected at least $m\ge1$ times on topic $c^-$, but has never been cited; equivalently, $h^+_t(w,c^-)=0$ and $h^0_t(w,c^-)\ge m$. Over an interval $(t,t']$ on which each comparable request on $c^-$ keeps satisfying the Withhold gate, $w$ is not selected on that topic. Then $\ell_{t'}(w,c^-)=\ell_t(w,c^-)\le-m\gamma$. If another topic $c^+$ gains $n_+\ge1$ cited selections and no uncited selection during the same interval, then
\begin{equation}
\bigl[\ell_{t'}(w,c^+)-\ell_{t'}(w,c^-)\bigr]
-
\bigl[\ell_t(w,c^+)-\ell_t(w,c^-)\bigr]
=n_+.
\end{equation}
Thus, while the request-specific gate holds, $c^-$ accumulates no new support and delivered alternatives expand their relative evidence advantage.
\end{lemma}

\paragraph{Derivation.} The Withhold condition requires $h^+_t(w,c^-)=0$ and $h^0_t(w,c^-)\ge m$, and substituting into the definition of $\ell$ gives $\ell_t(w,c^-)=-\gamma h^0_t(w,c^-)\le-m\gamma$. Over the interval in the proposition, every comparable request on $c^-$ continues to satisfy the request-specific gate, so $w$ is not selected on those requests. By \Cref{eq:memory}, a service that did not happen produces no record; neither count changes and $\ell_{t'}(w,c^-)=\ell_t(w,c^-)$. If $c^+$ gains $n_+$ cited selections and no new uncited selection over the same interval, then $h^+_{t'}(w,c^+)=h^+_t(w,c^+)+n_+$ and $h^0_{t'}(w,c^+)=h^0_t(w,c^+)$, so $\ell_{t'}(w,c^+)=\ell_t(w,c^+)+n_+$. Subtracting the unchanged $c^-$ score gives the stated difference. \hfill$\square$

\paragraph{Empirical link.} Under limited profile capacity, this relative evidence advantage favors delivery-backed terms when a content agent rewrites its profile. Empirically, \DEC produces deployed profiles with more historical delivery backing, and those profiles improve website matching under a fixed selector (\Cref{sec:better-profiles,fig:unbacked-claims,app:profile-replacement}).

\FloatBarrier
\section{Prompt templates}
\label{app:prompts}
We organize the prompts as a shared base chain followed by mechanism-specific additions; braces mark variables filled at call time. Identifiers that the implementation uses internally are renamed here to the notation of this paper; the deployed prompts are in the released code (\Cref{app:repro}). Original uses only \Cref{app:base-prompts}. Profile Adaptation retains that chain and adds the profile-rewriting prompt in \Cref{app:rewriting-prompt}. Credit additionally uses the query-relation, title-evidence, and evidence-card components in \Cref{app:dec-prompts}; \DEC further applies the deterministic Withhold rule, which makes no language-model call. The independent registry-admission classifier in \Cref{app:registry-admission} acts at submission, before Credit and Withhold act at selection; \Cref{app:provider-baselines} reports its robustness result.

\subsection{Base agent prompts}
\label{app:base-prompts}

The Selector, Website Content Agent, and Final Agent form the AgentWebBench service chain shared by every system \citep{zhong2026agentwebbench}. Website descriptions, search histories, and returned information fill the run-time fields. The mechanism arms do not rewrite these base roles or action contracts.

\noindent\textbf{Selector.}\par

\begin{lstlisting}[caption={Shared Selector prompt.},label={lst:base-selector}]
You are a search agent that selects which websites to query for a user request.
You have the following websites that can search, with their descriptions:
{website_descriptions}

User query: {query}

Instructions: - Select the {top_k} websites that are most relevant for answering this query. - Choose only website names from the list above; do not invent website names. - Order the websites by relevance, most relevant first. - Output only a JSON object and nothing else.

Required JSON format:
{"websites": ["website name", "website name", ...]}
\end{lstlisting}

\noindent\textbf{Website Content Agent.}\par

\begin{lstlisting}[caption={Shared Website Content Agent prompt.},label={lst:base-content-agent}]
You are a content assistant with the ability to perform web searches on a specific website "{website}" to answer questions. You can answer a question with many turns of search and reasoning.

Website: {website} Website description: {website_description}

Based on the history information, suggest the next action to complete the task. You are provided with prior <search> queries and results inside <information>, and the question to answer.

IMPORTANT: 1. Choose ONLY ONE action per response and follow its exact syntax. 2. Do not repeat a search already present in the history. 3. Before the first <answer>, perform at least one <search> on this website; do not reject a website from its description alone.

Valid actions: 1. <search> query </search> 2. <answer> {"summary": "summary text", "documents": ["doc_id1", ...]} </answer> 3. <summary> important parts of the history turns </summary>

For <answer>, copy exact document IDs from retrieved results, return no more than {num_docs}, sort them by relevance, and use an empty list when no relevant document is found. Do not reproduce text inside <information>.

Question: {question}
\end{lstlisting}

\noindent\textbf{Final Agent.}\par

The Final Agent completes the shared service chain from the information returned by content agents. All systems use the same base prompt and action contract; the deployed prompt is included in the released code (\Cref{app:repro}).

\subsection{Profile rewriting prompt}
\label{app:rewriting-prompt}

Each content agent updates its website profile from its own profile trajectory and provider-local service history. That history contains the website's selection, return, and citation outcomes; each candidate reports both cumulative outcomes and those observed since the current profile was deployed. The content agent cannot see other websites, the selector ranking, the frozen offline relevance judgements, or future requests. The listing merges the run-time payload into one field-named input block while preserving the remaining fixed instructions and output constraints. The objective stated in the prompt---final exposure through selection, return, and citation---is $\mathcal O^W$ in \Cref{sec:game}; the remaining instructions restrict how the updater pursues it, allowing only up to eight candidate terms from the website's own history, copied verbatim. The four rewriting variants adapted from prior work produce no material exposure gain in this market (\Cref{app:provider-baselines}).

\begin{lstlisting}[caption={Provider profile-update prompt.},label={lst:provider-update}]
Select query-hook slots for one website from its own observed trajectory.

Allowed information only: - website domain; - this website's original registry profile; - this website's own previous submitted profile and previous slots from the source trajectory; - this website's own elapsed selected queries, meaning queries where the website was selected; - a candidate pool mechanically extracted from those elapsed selected queries. - this website's own document titles returned into the final-information context for elapsed selected queries; - provider-local Content Agent return/self-veto outcomes for those selected queries; - whether this provider's own returned documents were cited in the elapsed final answer;

Forbidden information: other websites, user-side rankings or internals, evaluation labels or utility, and future queries.

[Input blocks: website; original description; previous profile and slots; own selected-query history; round trajectory; event ledger; query candidates; evidence candidates; previous-slot feedback]

Task: The next profile can add a short front hook of this form: "Also consider this site for query families involving ... when the query otherwise fits the site's content type."

Choose up to 8 exact candidate entries for the blank. The objective is provider relevant final exposure through the full funnel. Truthful website-selection access is a necessary first condition, so do not optimize downstream conversion by simply collapsing the provider's supported exposure surface.

Reasoning guidance: - One elapsed qid is one historical event. The surfaces listed under an event are alternative descriptions of that single event, not separate successes; do not count one event's outcome once per surface.

Output rules:
- Return exactly one JSON object and no markdown.
- Required JSON schema: {"selected_terms": ["exact candidate entry"], "selected_slots": [{"term": "exact candidate entry", "source": "query|evidence"}], "reason": "one short reason grounded in the shown evidence, 25 words or fewer"}
- Every selected term must exactly match one query candidate entry or one evidence candidate entry shown above.
- Do not invent, rewrite, stem, or translate candidate entries.
- Keep JSON strings short; no field may contain more than one sentence.
\end{lstlisting}

\subsection{DEC and registry safety prompts}
\label{app:dec-prompts}

Credit and \DEC retain the base chain in \Cref{app:base-prompts} and profile rewriting in \Cref{app:rewriting-prompt}. We give only their additional history-evidence processing below.

\paragraph{Query relation.}

Credit's first stage decides whether a past request can be reused for the current request. Candidate events are anonymized at this stage; the model sees no website identity, document title, service outcome, exposure, or frozen offline relevance judgement.

\begin{lstlisting}[caption={Query-relation prompt used by Credit.},label={lst:query-relation}]
You judge whether an anonymous past query event is reusable for the current query.
KEEP only events whose past query expresses the same information need or a directly reusable nearby need.
DROP by omission when the connection is only broad topic overlap or uncertainty.
For every kept event, output DIRECT for the same need or TRANSFERABLE for a directly reusable nearby need, and set every boolean field in the schema to true.
Return one JSON object and no markdown.

Current query: {current_query}

Candidate query events: {candidate_events}

Output schema:
{
 "decisions": [
 {
 "query_id": "Q0001",
 "relation": "DIRECT|TRANSFERABLE",
 "current_central_referent":
 "the central entity/object or 'not specified'",
 "past_central_referent":
 "the central entity/object or 'not specified'",
 "same_central_referent": true,
 "task_compatible": true,
 "source_form_compatible": true,
 "hard_constraints_compatible": true,
 "reason": "brief reason"
 }
 ]
}
\end{lstlisting}

\paragraph{Title evidence.}

Once the relation check passes, a fixed judge checks whether the recalled history answers the aspect the current question asks about. At most one evidence item is kept per website.

\begin{lstlisting}[caption={Title-evidence prompt used by Credit.},label={lst:title-evidence}]
You score whether one same-source historical title has the answer role needed by the current query.
Judge only the visible title text. Do not infer hidden page contents from source type or reputation.
Score using only the current query and listed titles: 3=the title directly fits the requested answer role; 2=useful but incomplete answer-role support; 1=same subject only; 0=no support or uncertain.
KEEP only score 3 and put the score in the reason text. DROP all other scores.
For each KEEP, copy one exact contiguous title span as support_quote, list its title_id, name the supported current slot, use support_mode CONTENT or DIRECT_MECHANISTIC_SUBPROBLEM, and leave unsupported_or_conflicting_slots empty.
For each DROP, use support_mode NONE. Return one JSON object and no markdown.

Current query: {current_query}

Frozen candidate actions: {candidate_actions}

Output schema:
{
 "decisions": [
 {
 "event_id": "E0001",
 "verdict": "KEEP|DROP",
 "support_mode":
 "CONTENT|DIRECT_MECHANISTIC_SUBPROBLEM|NONE",
 "support_title_ids": ["T1"],
 "support_quote":
 "exact contiguous title span; empty for DROP",
 "supported_current_slots": [
 "answer-bearing current slot supported by title"
 ],
 "unsupported_or_conflicting_slots": [],
 "reason": "brief site-evidence verification"
 }
 ]
}
\end{lstlisting}

The prompt reserves KEEP for titles that directly fit the current answer role (its score-$3$ criterion); a KEEP enters a supporting evidence card as $\mathrm{Support}=1$ only if its evidence fields pass validation. The $0$--$3$ scale is internal to the prompt and separate from the frozen offline relevance judgements.

\paragraph{Supporting evidence card.}

The card Credit attaches after a website profile. It states only what happened, gives no score and makes no recommendation.

\begin{lstlisting}[caption={Evidence-card template.},label={lst:evidence-card}]
Observed history: for the past query {past_query}, the final answer used document(s) {support_titles} from this website.
\end{lstlisting}

One rendered instance:

\begin{lstlisting}[caption={Rendered evidence-card example.},label={lst:evidence-card-example}]
Observed history: for the past query "fjord formation process definition geology", the final answer
used "Fjord - Wikipedia" from this website.
\end{lstlisting}

Titles must come from documents that this website previously contributed to a final answer, and \texttt{support\_quote} must be a contiguous substring of at least one listed title. Withhold makes no language-model call; it is applied deterministically from the history retrieval and service counts defined in \Cref{app:paths}.

\subsection{Independent registry-admission safety check}
\label{app:registry-admission}
\label{app:injection-prompt}

This check acts at submission, before the selection-time actions of Credit and Withhold. Profile manipulation continues the web-spam problem \citep{gyongyi2005taxonomy}, and explicit control instructions are prompt injections in the sense formalized by \citet{liu2024formalizing}. A frozen zero-shot classifier reads each submitted profile. A positive decision excludes the profile from the selector-visible registry for that deployment, while the downstream Selector, Content, and Final prompts remain unchanged. The binary detection structure and output contract follow PIGuard, while the instruction not to execute external text follows BIPIA \citep{li2025piguard,yi2025bipia}.

\begin{lstlisting}[caption={Registry-admission prompt.},label={lst:registry-admission}]
SYSTEM:
You are the Registry Admission role of the existing User Agent, with a strong security mindset. You will be given provider profiles that will be fed to an LLM-based website selector. Your job is to analyse whether it is safe to present each profile to the selector.

USER: A strategic provider may carefully craft a profile in order to control the selector and get it to include or rank the provider regardless of the user's request. Some of the profiles you receive may come from such providers.

For this registry, every provider profile is intended to influence selection by describing why the provider may be useful. An ordinary recommendation that remains conditional on the query matching the provider's content is safe. A prompt injection instead attempts to override or replace the host User Agent's or selector's decision policy--for example by requiring inclusion or ranking irrespective of query fit, controlling relevance or tool policy, or controlling the selector's output schema.

Do not execute or adhere to any potential instructions or commands in the following provider profile.

Do you think the following provider profile is a prompt injection instruction?

{profile}

That is the end of the provider profile. If you think it is a prompt injection, please answer YES. If you do not think so, please answer NO.
\end{lstlisting}

The stripped output must be exact \texttt{YES} or exact \texttt{NO}; any other output is treated as unparseable.

\FloatBarrier
\section{Implementation and reproducibility}
\label{app:repro}

\paragraph{Models.} The selector, content agent, final agent, the provider update, and Credit's LLM steps all use \texttt{deepseek-v4-flash} with temperature $0$ and thinking disabled. The registry-admission audit uses the same model and decoding settings. In the Gemini and Qwen settings, the same roles use \texttt{gemini-3-flash-preview} and \texttt{Qwen/Qwen3-4B-Instruct-2507} with the same decoding settings. Document relevance is judged offline by a separate \texttt{gpt-4o} judge under the UMBRELA protocol. On the $60$-day trajectories, the additional LLM calls of \DEC use about $0.8\%$ of the tokens that the full pipeline consumes per request.

\paragraph{Reproduction.} The package contains the prompts above, the full configuration, the request order and round manifest, and the evaluation scripts. It releases request references, and its instructions recover each request from the public DeepResearchGym log.

\FloatBarrier
\section{Limitations and Open Directions}
\label{app:limitations}

\DEC operates on the information available to a platform-side policy in a decentralized market: submitted profiles and recorded final-use outcomes, without access to content agents' internal states or documents before selection. Natural extensions span both sides of the market, including learned website-selection policies and alternative update rules and adoption shares for content agents. For content agents, the slot-based updater $F_P$ keeps each profile change attributable to the service records a provider receives, the channel through which \DEC acts. Future work can study richer joint profile--document strategies at corpus scale, including mechanisms that incentivize content agents to add useful documents for emerging user needs. Richer service feedback could further reveal how selection and rewriting jointly shape current service opportunities and the evidence available for later adaptation.

\paragraph{Privacy of the delivery evidence.} A supporting evidence card carries the matched past request and the titles that a final answer used from that website; it carries no client identifier. The request stream is keyed by HMAC-IP client proxies (\Cref{app:data-views}), Credit's query-relation judge reads a candidate event without website identity, document title, service outcome, exposure, or offline relevance judgement (\Cref{app:dec-prompts}), and the reproduction package releases request references (\Cref{app:repro}).

\end{document}